\documentclass[11pt,preprint,longabstract]{aastex}

\slugcomment{}

\shorttitle{SNLS3: Cosmological Analysis}
\shortauthors{Sullivan et al.}

\newcommand{\col}{\ensuremath{\mathcal{C}}}

\newcommand{\sigint}{\ensuremath{\sigma_{\mathrm{int}}}}
\newcommand{\chidof}{\ensuremath{\chi^2/\mathrm{DOF}}}
\newcommand{\ebmvmw}{\ensuremath{E_{B-V}^{\mathrm{mw}}}}

\newcommand{\wo}{\ensuremath{w_0}}
\newcommand{\wa}{\ensuremath{w_a}}

\newcommand{\mbmodel}{\ensuremath{m_{B}^{\mathrm{mod}}}}
\newcommand{\omatter}{\ensuremath{\Omega_{m}}}
\newcommand{\olambda}{\ensuremath{\Omega_{\Lambda}}}
\newcommand{\ok}{\ensuremath{\Omega_{k}}}
\newcommand{\ode}{\ensuremath{\Omega_{\mathrm{DE}}}}

\newcommand{\zhel}{\ensuremath{z_{\mathrm{hel}}}}
\newcommand{\zcmb}{\ensuremath{z_{\mathrm{cmb}}}}

\newcommand{\mB}{\ensuremath{m_{B}}}

\newcommand{\mstellar}{\ensuremath{\mathrm{M}_{\mathrm{stellar}}}}
\newcommand{\scriptm}{\ensuremath{\mathcal M_{B}}}

\defcitealias{2010A&A...523A...7G}{G10}
\defcitealias{2011ApJS..192....1C}{C11}

\begin{document}

\title{SNLS3: Constraints on Dark Energy Combining the Supernova Legacy Survey Three Year Data with Other Probes}

\author{
 M.~Sullivan\altaffilmark{1}, J.~Guy\altaffilmark{2},
 A.~Conley\altaffilmark{3,4}, N.~Regnault\altaffilmark{2},
 P.~Astier\altaffilmark{2}, C.~Balland\altaffilmark{2,5},
 S.~Basa\altaffilmark{6}, R.~G.~Carlberg\altaffilmark{3},
 D.~Fouchez\altaffilmark{7}, D.~Hardin\altaffilmark{2},
 I.~M.~Hook\altaffilmark{1,8}, D.~A.~Howell\altaffilmark{9,10},
 R.~Pain\altaffilmark{2}, N.~Palanque-Delabrouille\altaffilmark{11},
 K.~M.~Perrett\altaffilmark{3,12}, C.~J.~Pritchet\altaffilmark{13},
 J.~Rich\altaffilmark{11}, V.~Ruhlmann-Kleider\altaffilmark{11},
 D.~Balam\altaffilmark{13}, S.~Baumont\altaffilmark{14},
 R.~S.~Ellis\altaffilmark{15,1}, S.~Fabbro\altaffilmark{13},
 H.~K.~Fakhouri\altaffilmark{17}, N.~Fourmanoit\altaffilmark{2},
 S.~Gonz\'alez-Gait\'an\altaffilmark{3}, M.~L.~Graham\altaffilmark{9,10},
 M.~J.~Hudson\altaffilmark{18,19}, E.~Hsiao\altaffilmark{17},
 T.~Kronborg\altaffilmark{2}, C.~Lidman\altaffilmark{20},
 A.~M.~Mourao\altaffilmark{16}, J.~D.~Neill\altaffilmark{21},
 S.~Perlmutter\altaffilmark{17,22}, P.~Ripoche\altaffilmark{17,2},
 N.~Suzuki\altaffilmark{17}, E.~S.~Walker\altaffilmark{1,22}
}

\altaffiltext{1}{Department of Physics (Astrophysics), University of Oxford, Keble Road, Oxford, OX1 3RH, UK}
\altaffiltext{2}{LPNHE, Universit\'e Pierre et Marie Curie Paris 6, Universit\'e Paris Diderot Paris 7, CNRS-IN2P3, 4 place Jussieu, 75252 Paris Cedex 05, France}
\altaffiltext{3}{Department of Astronomy and Astrophysics, University of Toronto, 50 St. George Street, Toronto, ON M5S 3H4, Canada}
\altaffiltext{4}{Center for Astrophysics and Space Astronomy, University of Colorado, 593 UCB, Boulder, CO 80309-0593, USA}
\altaffiltext{5}{Universit\'e Paris 11, Orsay, F-91405, France}
\altaffiltext{6}{LAM, CNRS, BP8, P\^ole de l'\'Etoile Site de Ch\^ateau-Gombert,  38, rue Fr\'ed\'eric Joliot-Curie, 13388 Marseille Cedex 13, France}
\altaffiltext{7}{CPPM, Aix-Marseille Universit\'e, CNRS/IN2P3, Marseille, France}
\altaffiltext{8}{INAF - Osservatorio Astronomico di Roma, via Frascati 33, 00040 Monteporzio (RM), Italy}
\altaffiltext{9}{Las Cumbres Observatory Global Telescope Network, 6740 Cortona Dr., Suite 102, Goleta, CA 93117, USA}
\altaffiltext{10}{Department of Physics, University of California, Santa Barbra, Broida Hall, Mail Code 9530, Santa Barbara, CA 93106-9530, USA}
\altaffiltext{11}{CEA, Centre de Saclay, Irfu/SPP, F-91191 Gif-sur-Yvette, France}
\altaffiltext{12}{Network Information Operations, DRDC Ottawa, 3701 Carling Avenue, Ottawa, ON, K1A 0Z4, Canada}
\altaffiltext{13}{Department of Physics and Astronomy, University of Victoria, PO Box 3055 STN CSC, Victoria BC, V8T 1M8, Canada}
\altaffiltext{14}{LPSC, UJF, CNRS/IN2P3, INPG, 53 rue des Martyrs, 38026 Grenoble Cedex, France}
\altaffiltext{15}{Department of Astrophysics, California Institute of Technology, MS 105-24, Pasadena, CA 91125, USA}
\altaffiltext{16}{CENTRA-Centro Multidisciplinar de Astrof\'{\i}sica and Dep. F\'{\i}sica, IST, Av. Rovisco Pais, 1049-001 Lisboa, Portugal}
\altaffiltext{17}{LBNL, 1 Cyclotron Rd, Berkeley, CA 94720}
\altaffiltext{18}{Department of Physics and Astronomy, University of Waterloo, 200 University Avenue West, Waterloo, Ontario N2L 3G1, Canada}
\altaffiltext{19}{Perimeter Institute for Theoretical Physics, 31 Caroline St. N., Waterloo, ON, N2L 2Y5, Canada}
\altaffiltext{20}{Australian Astronomical Observatory, P.O.\ Box 296, Epping, NSW 1710, Australia}
\altaffiltext{21}{California Institute of Technology, 1200 E. California Blvd., Pasadena, CA 91125, USA}
\altaffiltext{22}{Department of Physics, University of California, Berkeley, 366 LeConte Hall MC 7300, Berkeley, CA 94720-7300, USA}
\altaffiltext{23}{Scuola Normale Superiore, Piazza dei Cavalieri 7, 56126 Pisa, Italy}

\email{sullivan@astro.ox.ac.uk}

\begin{abstract}

  We present observational constraints on the nature of dark energy
  using the Supernova Legacy Survey three year sample (SNLS3) of
  \citet{2010A&A...523A...7G} and \citet{2011ApJS..192....1C}.  We use
  the 472 SNe Ia in this sample, accounting for recently discovered
  correlations between SN~Ia luminosity and host galaxy properties,
  and include the effects of all identified systematic uncertainties
  directly in the cosmological fits. Combining the SNLS3 data with the
  full WMAP7 power spectrum, the Sloan Digital Sky Survey luminous red
  galaxy power spectrum, and a prior on the Hubble constant $H_0$ from
  SHOES, in a flat universe we find $\omatter=0.269\pm0.015$ and
  $w=-1.061^{+0.069}_{-0.068}$ (where the uncertainties include all
  statistical and SN Ia systematic errors) -- a 6.5\% measure of the
  dark energy equation-of-state parameter $w$. The statistical and
  systematic uncertainties are approximately equal, with the
  systematic uncertainties dominated by the photometric calibration of
  the SN~Ia fluxes -- without these calibration effects, systematics
  contribute only a $\sim2$\% error in $w$.  When relaxing the
  assumption of flatness, we find $\omatter=0.271\pm0.015$,
  $\ok=-0.002\pm0.006$, and $w=-1.069^{+0.091}_{-0.092}$.
  Parameterizing the time evolution of $w$ as $w(a)=\wo+\wa(1-a)$,
  gives $\wo=-0.905\pm0.196$, $\wa=-0.984 ^{+1.094}_{-1.097}$ in a
  flat universe. All of our results are consistent with a flat, $w=-1$
  universe.

  The size of the SNLS3 sample allows various tests to be performed
  with the SNe segregated according to their light curve and host
  galaxy properties.  We find that the cosmological constraints
  derived from these different sub-samples are consistent. There is
  evidence that the coefficient, $\beta$, relating SN~Ia luminosity
  and color, varies with host parameters at $>4\sigma$ significance
  (in addition to the known SN luminosity--host relation); however
  this has only a small effect on the cosmological results and is
  currently a sub-dominant systematic.

\end{abstract}


\keywords{cosmology: observations -- cosmological parameters -- supernovae: general -- surveys}

\section{Introduction}
\label{sec:introduction}

The discovery of the accelerating universe ranks as one of science's
landmark achievements in the 20th century.  Surveys exploiting distant
Type Ia supernovae (SNe~Ia) as standardizable candles
\citep{1998AJ....116.1009R,1999ApJ...517..565P} revealed the presence
of a ``dark energy'' that opposes gravity and accelerates the
expansion of the Universe.  When these SN observations are combined
with measures of large-scale structure
\citep{2005MNRAS.362..505C,2005ApJ...633..560E,2007ApJ...657...51P,2010MNRAS.404...60R}
and the cosmic microwave background \citep[CMB;
e.g.,][]{2002ApJ...564..559D,2003ApJS..148....1B,2011ApJS..192...16L},
this dark energy emerges as the dominant component of the Universe
responsible for 70--75\% of its energy density at the present epoch.

A compelling physical explanation of dark energy remains distant
despite a range of possibilities being postulated \citep[for reviews
see][]{2006IJMPD..15.1753C,2008ARA&A..46..385F}.  Astrophysical
measurements of the dark energy's equation-of-state parameter $w$ (the
ratio of its pressure to density, $p/\rho$), and its variation over
cosmic history, can help distinguish the possibilities.  The classical
``Cosmological Constant'' is equivalent to a vacuum energy density
with negative pressure, constant in time and space: $w=-1$.  The broad
family of ``quintessence'' models, a dynamical form of scalar energy
field, mostly predict $-1\le w<-\onethird$.  A measurement of
$w<-1$ would be a signal of even more exotic physics.

SNe~Ia remain, at present, the most direct and mature method of
probing this dark energy due to several decades of intensive study and
use in cosmology \citep[see the review of][]{2010arXiv1011.0441H}.
Thought to be the result of the thermonuclear destruction of an
accreting CO white dwarf star approaching the Chandrasekhar mass limit
\citep[e.g.,][]{2000ARA&A..38..191H}, they are standardizable candles
which explode with nearly the same brightness everywhere in the
Universe due to the uniformity of the triggering mass and hence the
available nuclear fuel. Their cosmological use exploits simple
empirical relations between their luminosity and other parameters.
Brighter SNe~Ia have slower, wider light curves
\citep{1993ApJ...413L.105P} and are typically bluer than their faster,
fainter counterparts \citep{1996ApJ...473...88R,1998A&A...331..815T}.
Detailed searches for a ``third parameter'' have recently shown that,
after applying these first two corrections, brighter SNe Ia reside in
more massive host galaxies
\citep{2010ApJ...715..743K,2010MNRAS.406..782S,2010ApJ...722..566L}.

The application of relations between SN~Ia light-curve-shape, color,
and host galaxy properties provides robust distance estimates which
allow SNe~Ia to be used to measure cosmological parameters.  When
combined together, recent SN~Ia samples
\citep{2006A&A...447...31A,2007ApJ...659...98R,2007ApJ...666..674M,2008AJ....136.2306H,2009ApJ...700..331H,2010AJ....139..519C}
provide measures of dark energy generally consistent with a
cosmological constant of $w=-1$ with statistical uncertainties in $w$
of around 5--7\%, and systematic uncertainties of 8--14\%, depending
on the method used and assumptions made
\citep{2009ApJS..185...32K,2009ApJ...700.1097H,2010ApJ...716..712A}.
These SN~Ia samples are now sufficiently large that our understanding
of systematic uncertainties has a direct impact on our measurement of
dark energy \citep[][]{2009ApJS..185...32K,2011ApJS..192....1C},
particularly when combining SNe Ia from many different surveys at
different observatories.

Systematic uncertainties which affect the cosmological analysis of
SNe~Ia arise from two broad sources. The first is experimental
systematics; for example, photometric calibration or survey selection
biases. Due to the correlations between SN~Ia magnitudes that these
uncertainties introduce, accounting for their effects in the
cosmological fits is particularly important.  Fortunately this is a
tractable problem as the uncertainties are understood in modern SN~Ia
samples and can be accurately modeled, albeit only after detailed and
painstaking work
\citep[][]{2009A&A...506..999R,2010A&A...523A...7G,2011ApJS..192....1C}.
The second source of systematics arises from an incomplete
understanding of their astrophysics (e.g., progenitor configuration,
mass transfer and explosion mechanism, etc.). The most pernicious
possibilities include evolution in SN~Ia properties with redshift
tracking changing metallicities/ages of the progenitor stars, and
varying dust extinction or color laws; the correct treatment of SN~Ia
color--luminosity relationships are particularly uncertain
\citep[e.g.,][]{2007ApJ...664L..13C}. The effects of these potential
systematics are more nebulous due to the difficulty in modeling SNe~Ia
explosions, but can be investigated empirically. Studies which compare
local SN~Ia spectra with those at high-redshift find a remarkable
degree of similarity across 2780\AA\ to 6000\AA\
\citep{2005AJ....130.2788H,2006AJ....131.1648B,2008ApJ...684...68F,2008ApJ...674...51E,2009A&A...507...85B,2011ApJ...727L..35C}
with only small differences in the strengths of some intermediate mass
element features \citep{2009ApJ...693L..76S}, consistent with a mildly
evolving mix in SN~Ia demographics with redshift expected from popular
SN~Ia delay-time distribution models
\citep[][]{2005A&A...433..807M,2006ApJ...648..868S,2006MNRAS.370..773M,2007ApJ...667L..37H}.
To date, no definitive evolutionary signature with redshift, which
would directly impact a cosmological analysis, has been located.

This paper presents the cosmological analysis of the three year
Supernova Legacy Survey SN~Ia sample (SNLS3). Our sample and methods
are presented over the course of several papers. The first,
\citet{2009A&A...506..999R}, deals with the photometric calibration of
the SNLS SN~Ia fluxes and associated systematic uncertainties,
including corrections for spatial non-uniformities in the SNLS
photometric imager. \citet[][hereafter G10]{2010A&A...523A...7G},
presents the light curves of the SNLS SNe~Ia themselves, together with
a comparison of SN light curve fitting techniques, color laws,
systematics and parameterizations.  \citet[][hereafter
C11]{2011ApJS..192....1C}, discusses systematic effects in the
cosmological analysis (including covariance matrices accounting for
correlations between the distances to different SNe), presents light
curve parameterizations of external SNe~Ia used in the analysis,
describes the various light curve quality cuts made to produce the
combined sample, and provides the cosmological constraints obtained
from the SN~Ia data alone. Other papers describe the SNLS selection
biases \citep{2010AJ....140..518P}, the SN~Ia host galaxy information
\citep{2010MNRAS.406..782S}, and the spectroscopic confirmation and
redshift measurements
\citep{2005ApJ...634.1190H,2008A&A...477..717B,2009A&A...507...85B,2011MNRAS.410.1262W}.
This paper performs a cosmological analysis combining the SN-only
analysis of \citetalias{2011ApJS..192....1C} with other external,
non-SNe constraints.

We use 242 well-sampled SNe~Ia over $0.08<z<1.06$ from the SNLS
together with a large literature sample: 123 SNe~Ia at low-redshift,
14 SNe~Ia at $z\ga0.8$ from the Hubble Space Telescope, and 93 SNe~Ia
at intermediate redshift from the first year of the Sloan Digital Sky
Survey-II SN search. We include the effects of identified systematic
uncertainties directly in our cosmological fitting analysis using an
approach outlined in detail in \citetalias{2011ApJS..192....1C}. This
allows our cosmological parameter uncertainties to include systematic
as well as statistical uncertainties, with covariances between
different SNe which influence the cosmological fits accounted for.
Examples of effects which cause such covariances include common
photometric zeropoints for different SNe, or selection effects for SNe
from the same survey. Appropriate covariance matrices allowing other
users of this combined data set to directly include systematic effects
in subsequent analyses can be found in
\citetalias{2011ApJS..192....1C}.

The advantages of the enlarged SNLS data set are multiple. Most
obviously, this represents a threefold increase in the SNLS sample
size compared to the first year SNLS cosmological analysis presented
in \citet[][hereafter A06]{2006A&A...447...31A}, and as such provides
a significant improvement in the statistical precision of the
cosmological constraints. Several improvements in survey strategy were
made following the first year of SNLS, including a more regular
observing cadence together with longer $z$-band exposures, important
for the highest redshift events.  Moreover, the enlarged data set
allows sources of potential astrophysical systematics to be examined
by dividing our SN~Ia sample according to properties of either the SN
(e.g., light curve width) or its environment
\citep{2010MNRAS.406..782S}. The increased size of the SNLS data set
has also enabled a better understanding of SN~Ia light curve and
spectral properties (particularly at $\lambda$$<$3600\AA\ in the
rest-frame) with a corresponding improvement in the methods for
estimating their distances
\citep{2007ApJ...663.1187H,2007A&A...466...11G,2008ApJ...674...51E,2008ApJ...681..482C},
and handling their colors \citepalias{2010A&A...523A...7G}.  The full
three years of the SNLS data also allow an improved photometric
calibration of the light curves and a more consistent understanding of
the experimental characteristics \citep{2009A&A...506..999R}.

A plan of the paper follows. $\S$~\ref{sec:data-overview} provides a
brief overview of the SN~Ia data, and $\S$~\ref{sec:cosm-fits-method}
describes our methodology for determining the cosmological parameters.
Our cosmological results are presented in
$\S$~\ref{sec:results}.  $\S$~\ref{sec:cosm-sub-sampl} discusses
cosmological fits to various sub-samples of our SN population designed
to assess possibilities of astrophysical biases within the SN~Ia
sample.  We summarize and conclude in $\S$~\ref{sec:conclusions}.

\section{Supernova Data and Methodology Overview}
\label{sec:data-overview}

We begin by briefly reviewing the SN~Ia datasets and the various
techniques that we use in the cosmological analysis. Full details of
all of our procedures can be found in \citetalias{2010A&A...523A...7G}
and \citetalias{2011ApJS..192....1C}, as well as
\citet{2007A&A...466...11G}, \citet{2008ApJ...681..482C},
\citet{2010AJ....140..518P}, and \citet{2010MNRAS.406..782S}.

\subsection{The SN~Ia samples}
\label{sec:sn-ia-samples}

Our SN~Ia samples are divided into two categories: those discovered
and confirmed by the Supernova Legacy Survey (SNLS), and those taken
from the literature which sample different redshift ranges to SNLS.
The SNLS uses data taken as part of the five-year Canada-France-Hawaii
Telescope Legacy Survey (CFHT-LS).  CFHT-LS is an optical imaging
survey, the deep component of which conducted repeat imaging of 4
fields every 3--4 nights in dark time with four filters, allowing the
construction of high-quality multi-color SN light curves
\citepalias{2010A&A...523A...7G}.  Spectroscopic follow-up is used to
confirm SN types and measure redshifts, critical in obtaining clean
samples of SNe~Ia as reliable photometric identification techniques
have yet to be developed, despite recent progress
\citep{2010PASP..122.1415K,bazin2011}.  Candidates were prioritized
following the procedure outlined in \citet{2006AJ....131..960S}. SNLS
benefited from large time allocations on 8--10m class telescopes --
$\sim1500$ hours over five years -- including the Gemini North and
South telescopes, the European Southern Observatory Very Large
Telescopes, and the Keck telescopes. Nearly all our SN spectra are
published\footnote{Spectra and light curves for the SNLS3 sample are
  available at the University of Toronto's Research Repository,
  T-Space: \url{https://tspace.library.utoronto.ca/snls}, as well as
  in the cited papers.}
\citep{2005ApJ...634.1190H,2008A&A...477..717B,2008ApJ...674...51E,2009A&A...507...85B,2011MNRAS.410.1262W}.
All spectra are analyzed and uniformly typed according to the
classification schemes of \citet{2005ApJ...634.1190H} and
\citet{2009A&A...507...85B}. Further information on all 242 SNLS SNe
Ia that we use, including light curve parameterizations, can be found
in \citetalias{2010A&A...523A...7G}.

The SNLS dataset is complemented with SNe Ia from the literature over
redshift ranges that the SNLS sample does not cover. We use 123 SNe~Ia
at low-redshift ($z\la0.08$) from a variety of sources
\citep[primarily][]{1996AJ....112.2408H,1999AJ....117..707R,2006AJ....131..527J,2009ApJ...700..331H,2010AJ....139..519C},
14 SNe~Ia at $z\ga0.8$ from the HST-discovered sample of
\citet{2007ApJ...659...98R}, and 93 SNe~Ia over $0.06\la z\la0.4$ from
the first year of the Sloan Digital Sky Survey-II SN search
\citep{2008AJ....136.2306H}.  Light curve parameterizations and other
data for these events, on the same relative system as that of
\citetalias{2010A&A...523A...7G}, can be found in
\citetalias{2011ApJS..192....1C}.

We also considered including other SN~Ia samples that, at least in
part, probe the same redshift range as SNLS. However, we do not do
this for several reasons (see also \citetalias{2011ApJS..192....1C}
for a discussion of these points).  SNLS is designed to control
systematics as much as is possible -- a single telescope survey with a
well-understood photometric response and calibration, using deep
exposures in filters that allow the same rest-frame colors to be
measured for most of the redshift range that it probes. There is also
significant published information on the host galaxies
\citep{2010MNRAS.406..782S}, essential for the cosmological analysis
and which is not available for other higher-redshift samples.
Finally, SNLS is by far the largest and best observed (i.e., highest
signal-to-noise for each event) SN~Ia sample over $0.3\la z\la1.0$.
Adding other SNe to this might lead to marginally improved
cosmological constraints from a purely statistical perspective, but
would certainly lead to a much more complex and uncertain analysis of
systematic uncertainties when combining data from many surveys
conducted at many telescopes.

All SNe are corrected for Galactic extinction, Malmquist and other
selection biases \citep{2010AJ....140..518P,2011ApJS..192....1C}, and
peculiar velocities (at low redshift).

\subsection{Light curve fitting and distance estimation}
\label{sec:light-curve-fitting}

We parameterize the SN~Ia light curves for distance estimation using
the combined results from updated versions of two independent light
curve fitters -- SiFTO \citep{2008ApJ...681..482C} and SALT2
\citep{2007A&A...466...11G}.  Both techniques provide an estimate of
the SN peak rest-frame $B$-band apparent magnitude at the epoch of
maximum light in that filter, a measure of the SN light curve shape,
and an estimate of the SN optical $B-V$ color (\col). SiFTO
parameterizes the light curve in terms of stretch ($s$), while SALT2
uses a related parameter $x_1$.  The two light curve fitters are
compared in \citetalias{2010A&A...523A...7G}, which also provides
details of the techniques used to average them into a single light
curve parameterization for subsequent distance estimation. The
distance estimation technique used, based on the combined SiFTO/SALT2
light curve parameters, is described in
$\S$~\ref{sec:cosm-fits-method}.

Other light curve fitting and distance estimation techniques are
available. In particular, the MLCS2k2 fitter
\citep{2007ApJ...659..122J} has been widely used in previous SN Ia
analyses \citep[e.g.][]{2009ApJ...700.1097H,2009ApJS..185...32K}, and
the use of MLCS2k2 versus SALT2 has led to significantly different
cosmological parameters in some cases. The two techniques are
mathematically equivalent (to first order -- see section 4.2.3 in
\citetalias{2010A&A...523A...7G}), and many of the apparent
differences can instead be traced back to the training data and priors
on SN color.  We do not use MLCS2k2 in the SNLS3 analysis for several
practical reasons, discussed at length in
\citetalias{2010A&A...523A...7G} and \citetalias{2011ApJS..192....1C},
and which we summarize here.

\begin{figure}
\plotone{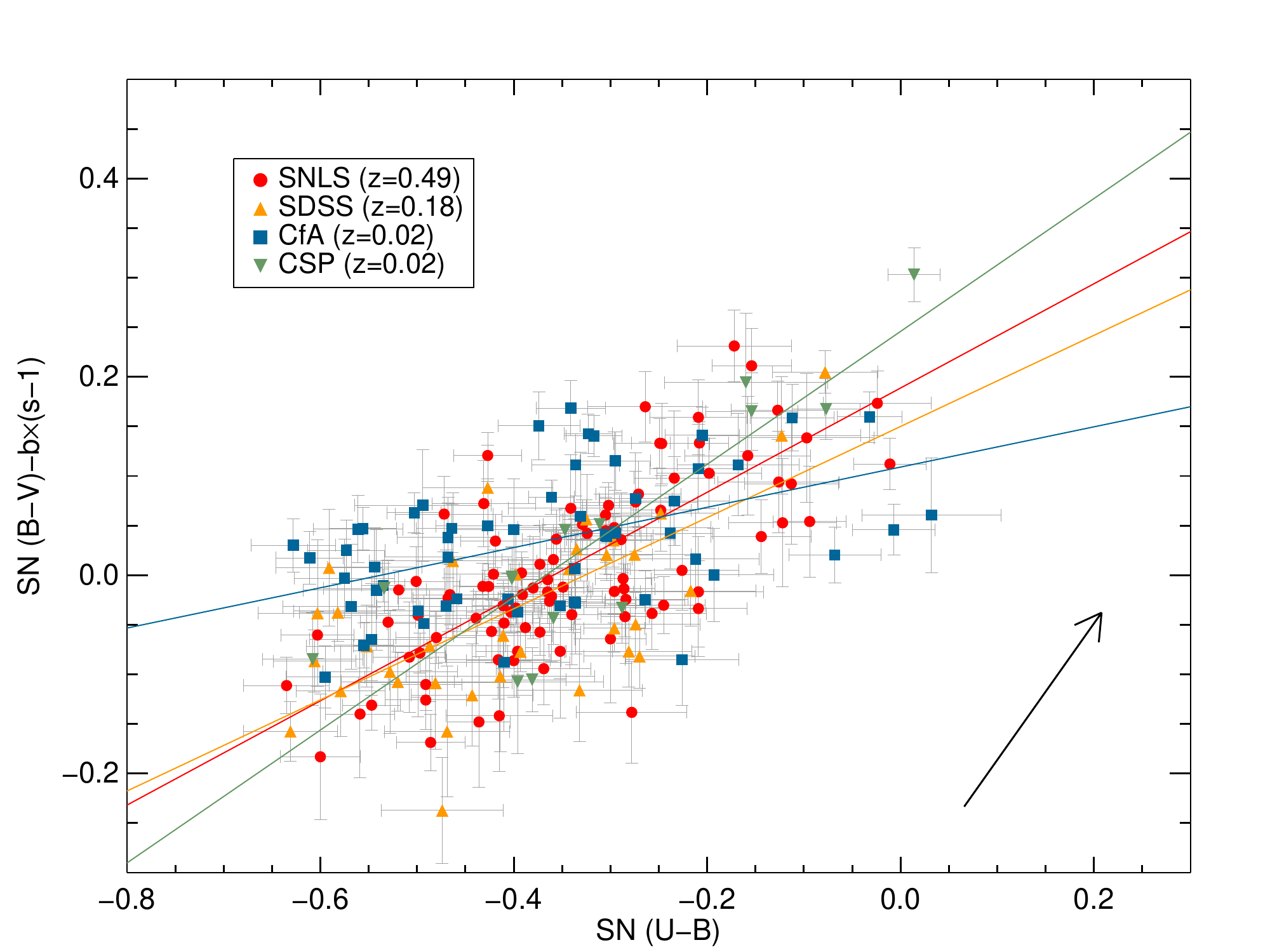}
\caption{SNe Ia from the SNLS3 sample in color--color space. The
  colors are measured at maximum-light in the rest-frame $B$-band. SNe
  are coded according to the sub-sample from which they are drawn with
  the mean redshifts shown, and the best-fitting SiFTO color--color
  law is over-plotted for each sample -- see
  \citet{2008ApJ...681..482C} for details of this fit. The arrow
  indicates the direction of the Milky Way extinction vector using a
  \citet{1989ApJ...345..245C} law. The CfA sample shows an offset
  from, and is statistically inconsistent with, the other
  samples.\label{fig:colcol}}
\end{figure}

There are apparent calibration problems with observer $U$-band SN Ia
data (which we do not use anywhere in our analysis) which MLCS2k2 is
reliant upon for its distance-estimation training. These include:
\begin{itemize}
\item Observer-frame $U$-band SN Ia data show more scatter around
  individual SN Ia light curve fits than can be accounted for by the
  published observational uncertainites. This large intrinsic
  scatter is not seen in SNLS and SDSS-SN SN observations
  transformed into rest-frame $U$-band.
\item In $U-B$ versus $B-V$ color-color space, the SNe with
  observer-frame $U$-band data show a systematic offset compared with
  the other SN samples (see Fig.~\ref{fig:colcol}),
\item There is significant tension between SNe with observer-frame
  $U$-band data, and those without, in the cosmological fits. This
  tension disappears if this $U$-band data is removed from the light
  curve fits for the low-$z$ SNe.  
\end{itemize}
A sample of low-$z$ SNe from the Carnegie Supernova Project
\citep{2010AJ....139..519C} with observer-frame $u'$ data show none of
these three problems above.  That is, the properties of the low-$z$ SN
Ia $U$-band data are inconsistent with SNLS at $z\sim0.5$, with
SDSS-SN at $z\sim0.25$, and with the CSP sample at $z\sim0.02$.
  
Various possibilities for the origin of this $U$-band anomaly are
outlined in \citetalias{2011ApJS..192....1C} section 2.6 \citep[see
also][]{2009ApJS..185...32K}. For this to be an evolutionary or
astrophysical effect, any evolution must be astonishingly sudden
(i.e., turning on at $z\sim0.25$ and then not evolving any further out
to $z=1$) -- yet somehow only effect one of the two sources of low-$z$
data (those with $U$, instead of $u'$, data). It must also somehow not
manifest itself in maximum-light spectral comparisons between low-$z$
and high-$z$
\citep[e.g.,][]{2008ApJ...674...51E,2008ApJ...684...68F,2011ApJ...727L..35C}.
Note that the small, and not very significant, evolution that is seen
in \citet{2011ApJ...727L..35C} is in the opposite sense to that
implied by the $U$-band photometry problem.
  
The overwhelming likelihood is that this is a problem with the
notoriously difficult calibration of the observer-frame $U$-band data
-- there is no evidence that it is an astrophysical effect, and
significant evidence that it is not.  Note that an MLCS2k2 trained
without the $U$-band data severely impacts the science that can be done
-- as a distance estimator, MLCS2k2 requires $z\lesssim0.06$ SNe Ia
for the training, and therefore cannot be supplemented with SNLS/SDSS
SN data sampling the rest-frame $U$-band as is the case with SALT2 and
SiFTO.

Even with a version of MLCS2k2 not trained using the current $U$-band
data, additional problems remain. These are discussed at length in
\citetalias{2010A&A...523A...7G} in their section 4.2.  In that
analysis, the authors noted that MLSC2k2 requires the use of priors
that color variation in SNe Ia is caused by dust extinction, which can
introduce additional biases into the estimated distances -- there is
no evidence that color variation in SNe Ia is caused purely by dust
extinction, and significant evidence that intrinsic SN properties make
the story more complex
\citep[e.g.,][]{2007ApJ...664L..13C,2009Natur.460..869K,2011ApJ...729...55F,2011A&A...529L...4C,2011ApJ...734...42N,2011MNRAS.413.3075M}.
Additionally, a \citet{1989ApJ...345..245C}-like color-variation law
for the SNe is assumed, which again is not supported by the data
\citep{2007A&A...466...11G,2010A&A...523A...7G}.

Disentangling the effect of intrinsic color variation from color
variation due to dust would require additional parameters in MLCS2k2,
SALT2 or SiFTO which currently do not exist -- at present no light
curve fitter correctly disentangles the two effects. In particular, a
weakness of the approach used in this paper (see
$\S$~\ref{sec:cosm-fits-method}) is to conflate intrinsic and dust
effects into a single parameter, $\beta$, during the distance
estimation.  However, the tests that we are able to perform -- for
example examining the evolution in this $\beta$ parameter with
redshift (figure 14 in C10) -- show no evidence for any significant
systematic effects in the SALT2/SiFTO distance estimation method.  The
case of maximal evolution in $\beta$ supported by our data are
included in our error budget (see $\S$~\ref{sec:syst-uncert}).

In conclusion, the differences reported in \citet{2009ApJS..185...32K}
are now understood and can be traced back to a combination of the
$U$-band calibration and color priors -- and should not, therefore, be
considered as systematics. These above issues were a significant
motivation for us to develop two independent light curve fitters
within SNLS that differ conceptually in the way that color is handled
\citep{2007A&A...466...11G,2008ApJ...681..482C,2010A&A...523A...7G}.
Differences in the light curves fits from these two codes are carried
through as an uncertainty in our analysis -- so our final quoted
errors on the cosmological parameters fully include this effect.

\subsection{SN Selection}
\label{sec:sn-selection}

We apply various selection cuts to the SN Ia samples designed to
ensure an adequate wavelength and phase coverage in the light curve
fits.  These are described in detail in
\citetalias{2010A&A...523A...7G} and \citetalias{2011ApJS..192....1C},
and essentially ensure that an accurate light curve width, rest-frame
color, and peak rest-frame $B$-magnitude can be measured.
Additionally, each SN must be spectroscopically confirmed
\citep[see][for a discussion of our spectroscopic classification
criteria]{2005ApJ...634.1190H}, have a minimum CMB-frame redshift
(\zcmb) of 0.010 (after peculiar velocity correction), be
spectroscopically normal, have a Galactic extinction of $\ebmvmw<0.2$,
and be of normal stretch ($0.7<s<1.3$) and color ($-0.25<\col<0.25$).
We also remove 6 outliers on the Hubble diagram -- see
\citetalias{2011ApJS..192....1C} for details.

\subsection{Host galaxy parameterizations}
\label{sec:host-galaxy-param}

Recent analyses have found correlations between SN~Ia luminosity and
their host galaxies, even after the well-known fainter--faster and
fainter--redder corrections have been made
\citep{2010ApJ...715..743K,2010MNRAS.406..782S,2010ApJ...722..566L}.
At the present time, it is not clear from an astrophysical perspective
which host galaxy parameter is the optimal choice to use (as the third
parameter) in the analysis. The observed effects could be due to
progenitor metallicity \citep[e.g.,][]{2009Natur.460..869K},
progenitor age \citep[e.g.,][]{2010ApJ...719L...5K}, or possibly some
other parameter.  Empirically, trends in SN~Ia luminosity are seen
with a variety of derived host galaxy parameters, including stellar
mass (\mstellar), star formation rate (SFR), and inferred stellar age.
However many of these parameters are strongly correlated when derived
from broad-band photometry available for the host galaxies.

We use the host galaxy stellar mass as the third variable in our
cosmological fitting. This has the advantage of being easiest to
determine from sometimes limited host data, and shows significant
trends with the SN~Ia luminosities. We derive the host information by
following the method in \citet{2010MNRAS.406..782S}, fitting the
broad-band spectral energy distribution (SED) of the host galaxies
using simple galaxy population synthesis models
\citep{2002A&A...386..446L}. The host galaxy information can be found
in \citetalias{2010A&A...523A...7G} and
\citetalias{2011ApJS..192....1C} for all the SNe used in our analysis.

\subsection{Systematic uncertainties}
\label{sec:syst-uncert}

We consider a variety of systematic uncertainties in our analysis,
discussed in detail in \citetalias{2011ApJS..192....1C}. Details of
the construction of the covariance matrices that encode this
information can be found in that paper. For each systematic we
estimate its size and adjust that variable in the light curve fits.
All the light curves are refit, including a re-training of the light
curve model where required, and the derived light curve parameters
(\mB, $s$, \col) compared for each SN with and without the inclusion
of the systematic.  These differences are converted into a covariance
matrix following \citetalias{2011ApJS..192....1C}.

\section{Cosmological fit methodology}
\label{sec:cosm-fits-method}

Having summarized the main features of our dataset, we now turn to the
cosmological analysis.  We write the $\chi^2$ as
\begin{equation}
\label{eq:cosmochi2}
\chi^2=\sum_{\mathrm{SNe}}\frac{\left(\mB - \mbmodel\right)^2}{\sigma_{\mathrm{stat}}^2+\sigint^2}
\end{equation}
where we have omitted the covariance error matrix for clarity.
$\sigma_{\mathrm{stat}}$ is the identified statistical error and
includes uncertainties in both \mB\ and \mbmodel, \sigint\
parameterizes the intrinsic dispersion of each SN sample (see below),
and the sum is over the SNe~Ia entering the fit. \mB\ are the
maximum-light SN rest-frame $B$-band apparent magnitudes and \mbmodel\
are the model $B$-band magnitudes for each SN given by
\begin{eqnarray}
\label{eq:mtheory}
\mbmodel=5\log_{10}{\mathcal D_L}\left(\zhel,\zcmb,w,\omatter,\ode,\ok\right)\\
\nonumber - \alpha\left(s-1\right) + \beta\col + {\mathcal M_{B}}
\end{eqnarray}
where $w$ is the equation of state parameter of dark energy, \omatter\
and \ode\ are the fractional energy densities of matter and dark
energy (for $w=-1$, $\ode\equiv\olambda$), \ok\ is the curvature
constant, and $\alpha$ and $\beta$ parameterize the $s$ and
\col--luminosity relationships. Any linear variation between SN
intrinsic color and $s$ will be absorbed into the
$\alpha$ term.  \zhel\ is the heliocentric redshift used in the light
curve fits. ${\mathcal D_L}$ is the $c/H_0$ reduced luminosity
distance with the $c/H_0$ factor absorbed into ${\mathcal M_{B}}$
(here $c$ is the speed of light and $H_0$ the Hubble constant).
Explicitly, ${\mathcal M_{B}}=M_B+5\log_{10}(c/H_0)+25$, where $M_B$
is the rest-frame absolute magnitude of a SN~Ia in the $B$-band.
Neither $H_0$ nor $M_B$ are assumed during the fitting process.

We allow ${\mathcal M_{B}}$ to vary as a function of host galaxy
stellar mass (\mstellar) to account for relations between SN~Ia
brightness and host properties that are not corrected for via the
standard $s$ and \col--luminosity relations following
\citet{2010MNRAS.406..782S}. Explicitly, we fit for ${\mathcal
  M_{B}^1}$ in galaxies with $\mstellar\leq10^{10}$\,M$_{\odot}$, and
${\mathcal M_{B}^2}$ when $\mstellar>10^{10}$\,M$_{\odot}$. We could,
of course, allow the other nuisance parameters $\alpha$ and $\beta$ to
vary according to host type -- we discuss this further in
$\S$~\ref{sec:cosm-sub-sampl}.

The statistical errors affecting each SN include the statistical error
in \mB\ from the light curve fit, the statistical error in \mbmodel\
(essentially $\alpha\sigma_{s}$ and $\beta\sigma_{\col}$), a peculiar
velocity error of 150 km\,s$^{-1}$ after correction for a local bulk
flow model, the error in \zhel\ projected into magnitude space, a 10\%
uncertainty from Milky Way extinction corrections
\citep{1998ApJ...500..525S}, a random scatter due to gravitational
lensing following \citet{2010MNRAS.405..535J} of
$\sigma_{\mathrm{lens}}=0.055z$, and the covariances between $s$,
\col\ and \mB\ for an individual SN (these parameters are correlated
as they are determined from the same light curve data).
$\sigma_{\mathrm{stat}}$ is updated during the fits as $\alpha$ and
$\beta$ are altered. The \sigint\ term parameterizes the extra
dispersion in \mB\ required to give a $\chi^2$ per degree of freedom
(DOF) of one in the cosmological fits
\citep[e.g.,][]{1999ApJ...517..565P}. This ``intrinsic'' dispersion
arises from unidentified sources of error in our analysis, as well as
the imperfect nature of SNe~Ia as standard candles. \sigint\ may also
include contributions from unidentified experimental errors and survey
selection effects, and there is no \textit{a priori} reason for
\sigint\ to be the same from SN sample to SN sample; we allow a
different \sigint\ for each sample, the values for which can be found
in table~4 of \citetalias{2011ApJS..192....1C}. These values are not
varied in the fits, but the values are fixed to give a $\chidof=1$ for
each sample in the SN only cosmological fits of
\citet{2011ApJS..192....1C}. Note that more sophisticated statistical
techniques for treating \sigint\ and its uncertainty have been
proposed \citep{2011arXiv1102.3237M}.

To include systematic errors we generalize eqn.~(\ref{eq:cosmochi2})
by constructing a covariance matrix $\mathbf{C}$ to replace the
$\sigma$ terms. $\mathbf{C}$ is the combination of a systematics
covariance matrix $\mathbf{C}_{\mathrm{syst}}$ and two covariance
matrices containing statistical uncertainties:
$\mathbf{C}_{\mathrm{stat}}$ which contains statistical errors from
the SN model used in the light curve fit and which are therefore
correlated between SNe, and $\mathbf{D}_{\mathrm{stat}}$, a purely
diagonal covariance matrix generated from the statistical errors
described above. We include both $\mathbf{C}_{\mathrm{stat}}$ and
$\mathbf{D}_{\mathrm{stat}}$ when performing fits based only on
statistical errors.

We then minimize the $\chi^2$ according to
\begin{equation}
\label{eq:cosmochi2withsys}
\chi^2=\sum_{N}\left(\vec{\mathbf{\mB}} - \vec{\mathbf{\mbmodel}}\right)^T\mathbf{C}^{-1}\left(\vec{\mathbf{\mB}} - \vec{\mathbf{\mbmodel}}\right)
\end{equation}
\noindent
This methodology allows the quoted uncertainties on the fit parameters
to directly include systematic errors, as well as correctly accounting
for systematic and statistical uncertainties which induce correlations
between different SNe and thus alter the position of the best-fit
cosmological model.

\subsection{Fitting techniques}
\label{sec:fitting-techniques}

We use three approaches\footnote{All the computer programs and code
  referred to in this paper are available at
  \url{https://tspace.library.utoronto.ca/snls}, along with the SN~Ia
  light curves, spectra, light-curve parameters, covariance matrices,
  and some of the {\tt CosmoMC} chains.} to perform our cosmological
fits. For relatively simple cosmological fits involving a small number
of parameters, we use a grid technique that computes the $\chi^2$ of
eqn.~(\ref{eq:cosmochi2withsys}) at every point converting into a
probability via $P\propto \exp\left(-\frac{1}{2}\chi^2\right)$, with
the proportionality set by normalizing over the grid.  The ``nuisance
parameters'' $\alpha$, $\beta$ and ${\mathcal M_{B}}$ are marginalized
over when generating confidence contours in the parameters of
interest, and we report the expectation value of the marginalized
parameters. Due to the (relatively) fast run-time, and the contour
visualization, this fitting technique is particularly well-suited to
analyzing the magnitude of the individual sources of systematic
uncertainty in our analysis, which would be impractical with more
complex and slower fitting approaches.

The second approach is a $\chi^2$ minimization routine which simply
reports the best-fit.  The results of this technique should be close
to the reported values from the grid marginalization, but should not
be expected to agree exactly, and we provide both. Note that the
\sigint\ calculated by \citetalias{2011ApJS..192....1C} is performed
for the marginalization approach fits -- when these \sigint\ are used
in the $\chi^2$ minimization fits, a $\chidof<1$ should be expected as
the best-fit parameters from the marginalization fits will not lie at
a minimum in $\chi^2$.

The third approach is the \texttt{CosmoMC} program
\citep{2002PhRvD..66j3511L}, which uses a Markov-Chain Monte Carlo
technique to explore cosmological parameter space.  We use this
approach for our main cosmological results. We made the following
modifications to the May 2010 version of \texttt{CosmoMC} package to
handle SNLS3 data: first, we properly marginalize over the SN nuisance
parameters $\alpha$ and $\beta$ rather than holding them fixed;
second, we keep track of the difference between heliocentric and CMB
frame redshifts, which enter into the luminosity distance differently,
important for some of the lowest-$z$ SNe; third, we have added the
ability to fit for the host-dependence of SN~Ia absolute luminosities
as described in \citet{2010MNRAS.406..782S}.

The first item above is handled most efficiently by explicitly fitting
for $\alpha$ and $\beta$ along with the cosmological parameters, as
internally marginalizing over their values is computationally more
expensive for the SNLS3 sample.  The consequences of incorrectly
holding the nuisance variables fixed, or of simply substituting the
values that minimize the $\chi^2$, both true of the default
\texttt{CosmoMC} SN~Ia implementation, are discussed in section~4.6
and Appendix~C of \citetalias{2011ApJS..192....1C}, as well as
$\S$~\ref{sec:results} of this paper. In brief, this simplified
approach leads to both underestimated uncertainties, and biased
parameter estimates, due to small correlations between $\alpha$,
$\beta$, and the cosmological parameters.

For \texttt{CosmoMC} fits where we allow for a time-varying dark
energy equation of state ($w(a)=\wo+\wa(1-a)$, where $a$ is the scale
factor), we follow the prescription of \citet{2008PhRvD..78h7303F}.
Further, in the \texttt{CosmoMC} fits we do not consider massive
neutrinos, and assume a simple power-law primordial power spectrum
(i.e., we neglect tensor modes, and any running of the scalar spectral
index).

\subsection{External, non-SN datasets}
\label{sec:external-datasets}

We include several external non-SN datasets in our fits. For the grid
marginalization and $\chi^2$ minimization fits, we use two external
constraints.  The first is the Sloan Digital Sky Survey (SDSS) Data
Release 7 (DR7) Baryon Acoustic Oscillations (BAO) measurements of
\citet{2010MNRAS.401.2148P}. This is a Gaussian prior on the distance
ratios $r_s(z_d)/D_V(z)$ at $z=0.2$ and $z=0.35$, where $r_s(z_d)$ is
the comoving sound horizon at the baryon drag epoch, and $D_V(z)$ is a
spherically averaged effective distance measure given by
$D_V(z)=[(1+z)^2D^2_A(z)cz/H(z)]^{1/3}$ \citep{2005ApJ...633..560E}
where $D_A(z)$ is the proper angular diameter distance. The second is
a prior based on the Wilkinson Microwave Anisotropy Probe 7-year
(WMAP7) ``shift'' parameter $R$ \citep{1997MNRAS.291L..33B}, the
``acoustic scale'' $l_a$, and the decoupling redshift $z_{\ast}$, as
defined in \citet{2011ApJS..192...18K}, following the prescription of
\citet{2009ApJS..180..330K}. This prior includes most of the power of
the CMB data for measuring dark energy
\citep[e.g.][]{2007PhRvD..76j3533W}.

For our main cosmological fits, with the \texttt{CosmoMC} program, we
use different external constraints: the power spectrum of luminous red
galaxies (LRGs) in the SDSS DR7 \citep{2010MNRAS.404...60R} in place
of the BAO constraints, the full WMAP7 CMB power spectrum
\citep{2011ApJS..192...16L} in place of the shift parameters, and a
prior on $H_0$ from the SHOES (Supernovae and $H_0$ for the Equation
of State) program \citep{2009ApJ...699..539R,2011ApJ...730..119R}.
This Gaussian $H_0$ prior, $H_0=73.8\pm2.4$ km\,s$^{-1}$\,Mpc$^{-1}$,
makes use in its construction of many of the low-redshift $z<0.1$ SNe
Ia used in this paper. Their absolute magnitudes are calibrated
directly using Cepheid variables in eight local SN~Ia host galaxies
\citep{2011ApJ...730..119R}, the Cepheids themselves calibrated using
different techniques: the geometric maser distance to the galaxy NGC
4258, trigonometric parallax distances for Milky Way Cepheids, and
eclipsing binary distances for Cepheids in the Large Magellanic Cloud.
The $H_0$ prior was derived with SN~Ia parameters from the MLCS2k2
distance estimator \citep{2007ApJ...659..122J} given in
\citet{2009ApJ...700.1097H}. In principle, for complete consistency
with this work, we would use the combined SALT2/SiFTO fits for the
same SN events to re-estimate $H_0$ in a consistent way. However, the
maser distance used in \citet{2011ApJ...730..119R} is not published,
so we defer this exercise to a future analysis. However, we note that
the uncertainty in $H_0$ given by Riess et al. does include an
allowance for the systematic of using SALT2 in place of MLCS2k2 (they
quote an increase in $H_0$ of 1.0 km\,s$^{-1}$\,Mpc$^{-1}$ with
SALT2), so most of the systematic difference is likely already
included in our error budget.

\section{Results}
\label{sec:results}

We begin by assessing the magnitude of the various systematic
uncertainties in our analysis. For this, we use a simple cosmological
model -- a flat cosmology with a constant $w$ -- and the grid
marginalization approach ($\S$~\ref{sec:cosm-fits-method}).

We then present our main cosmological results. We investigate a
non-flat, $w=-1$ cosmology (fitting for \omatter\ and \olambda), a
flat, constant $w$ cosmology (fitting for \omatter\ and $w$), a
non-flat cosmology with $w$ free (fitting for $w$, \omatter\ and \ok),
and a cosmology where $w(a)$ is allowed to vary via a simple linear
parameterization $w(a)=\wo+\wa(1-a)\equiv\wo+\wa z/(1+z)$
\citep[e.g.,][]{2001IJMPD..10..213C,2003PhRvL..90i1301L}, fitting for
\omatter, \wo, and \wa. We always fit for $\alpha$, $\beta$, and
${\mathcal M_{B}}$.

\begin{figure}
\includegraphics[width=0.49\textwidth]{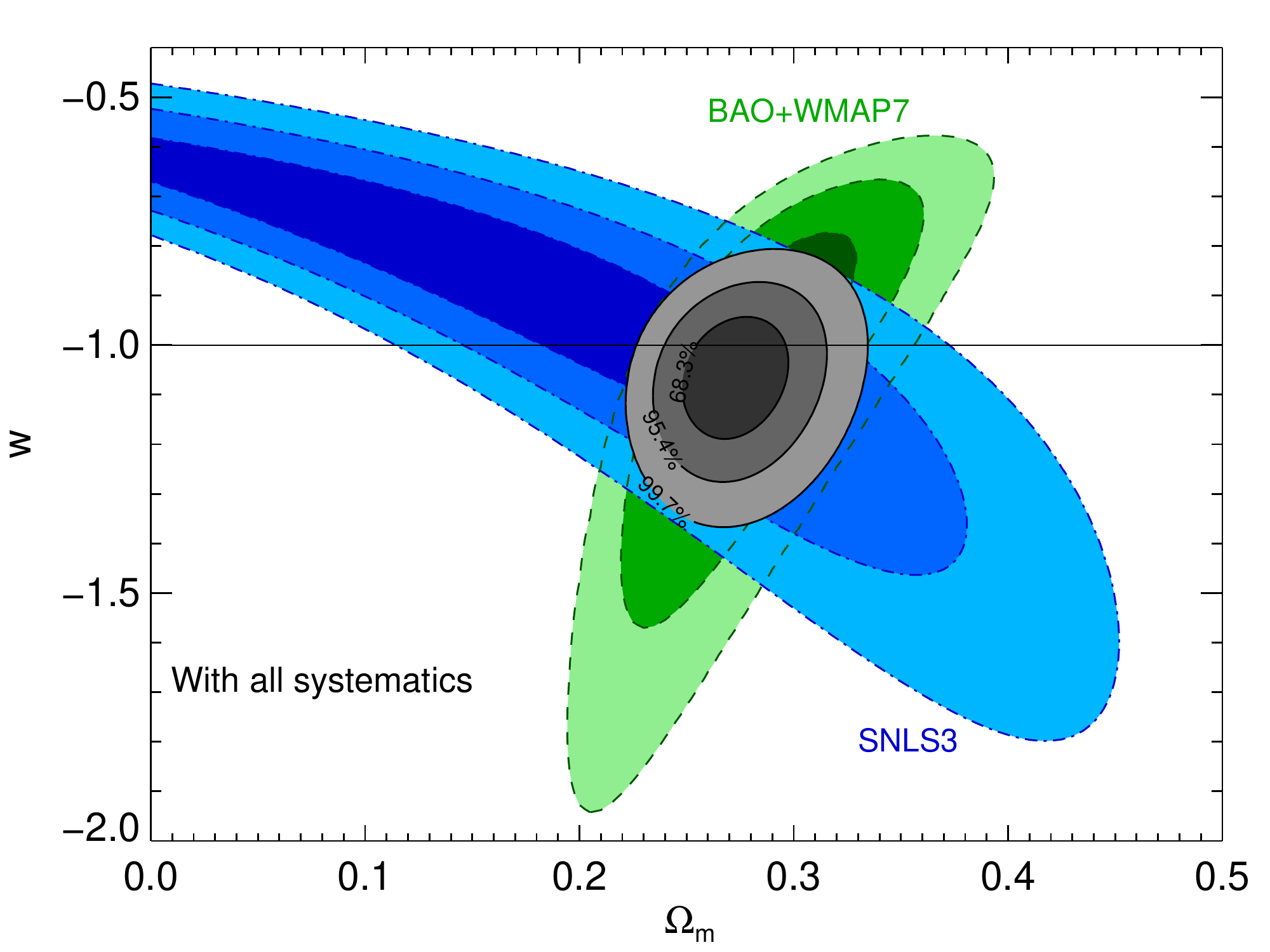}\includegraphics[width=0.49\textwidth]{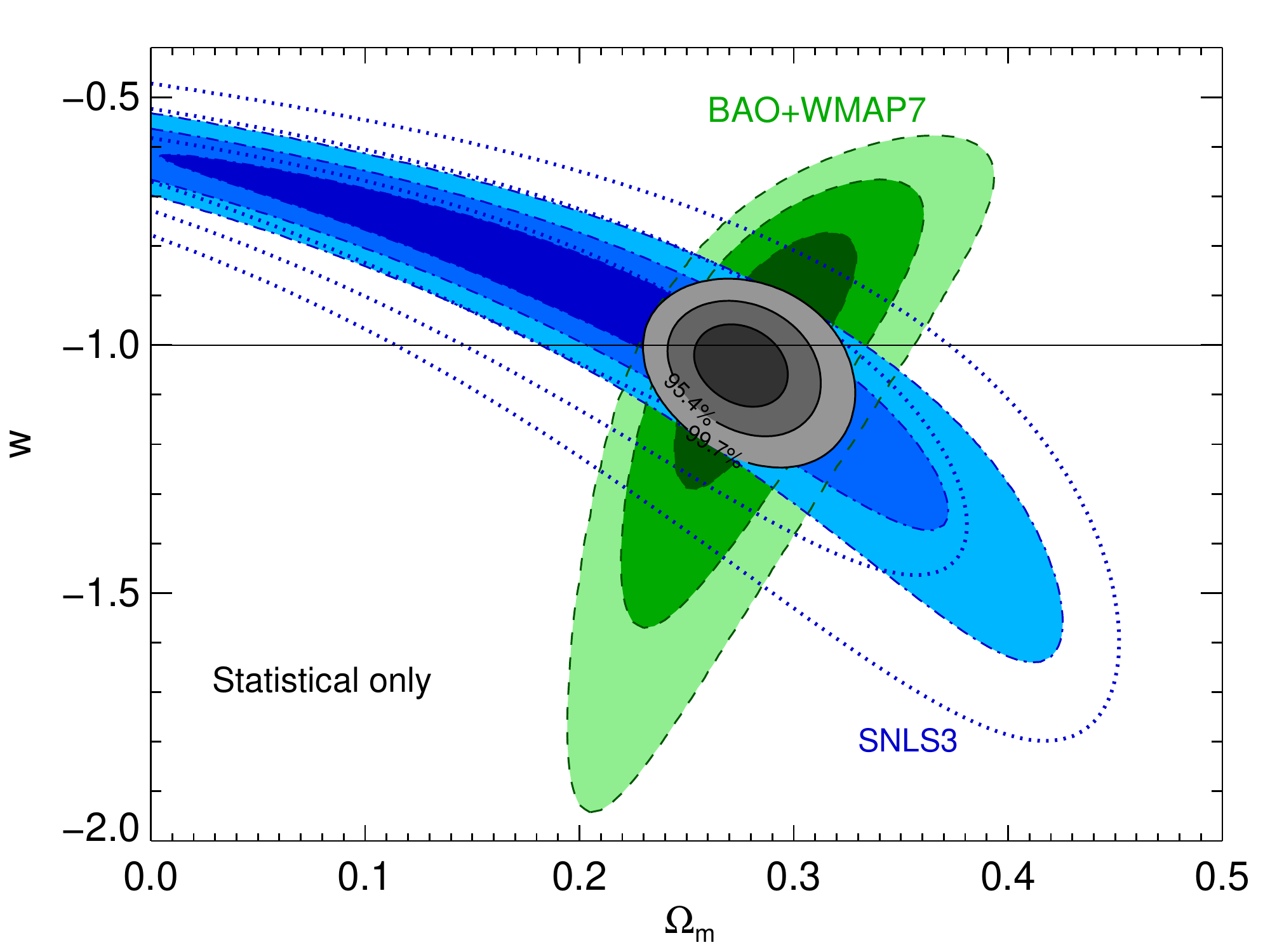}
\includegraphics[width=0.49\textwidth]{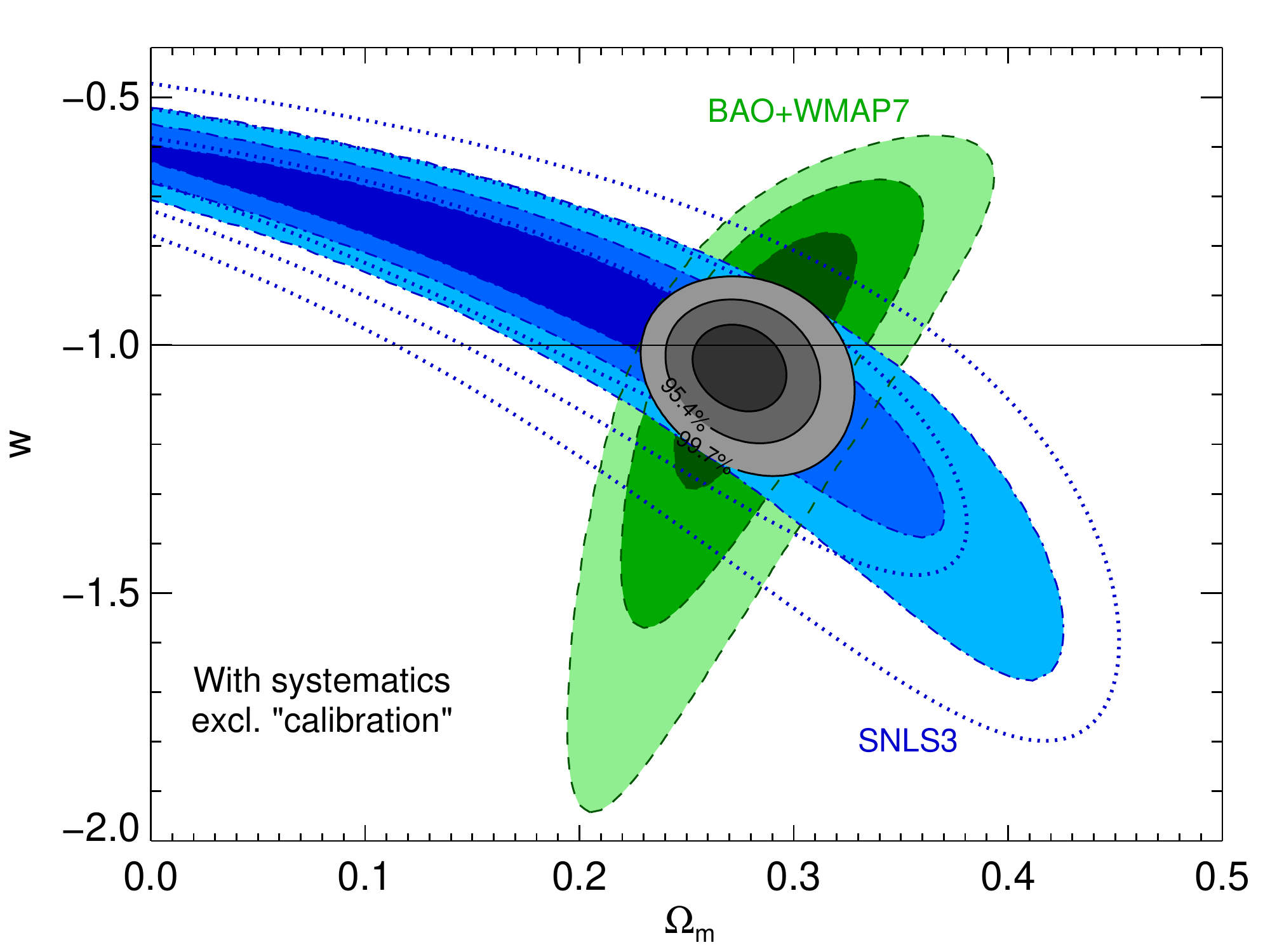}\includegraphics[width=0.49\textwidth]{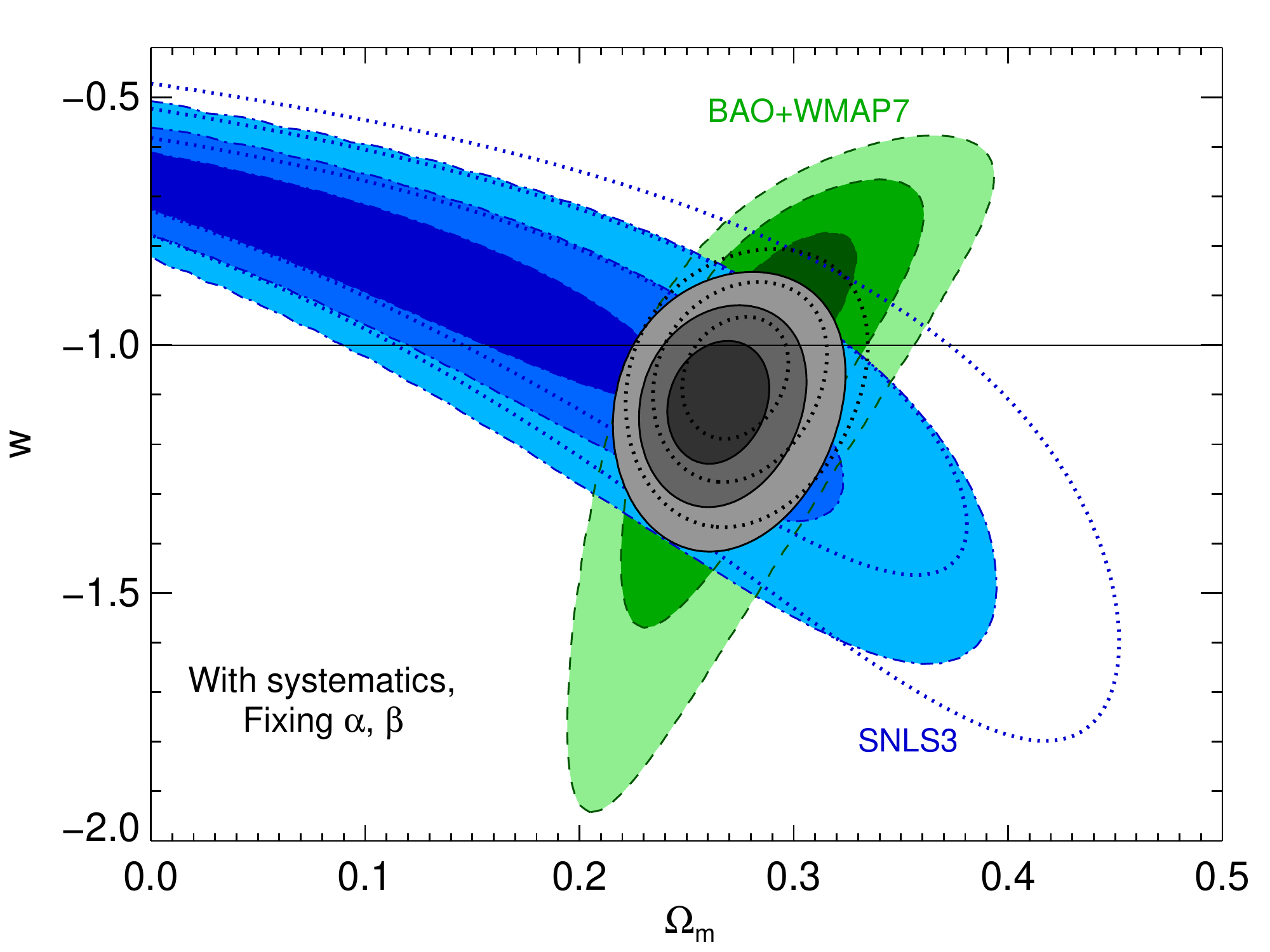}
\caption{Confidence contours in the cosmological parameters \omatter\
  and $w$ arising from fits to the combined SN~Ia sample using the
  marginalization fitting approach, illustrating various systematic
  effects in the cosmological fits. In all panels, the SNLS3 SN~Ia
  contours are shown in blue, and combined BAO/WMAP7 constraints
  \citep{2010MNRAS.401.2148P,2011ApJS..192...18K} in green. The
  combined constraints are shown in grey. The contours enclose 68.3\%,
  95.4\% and 99.7\% of the probability, and the horizontal line shows
  the value of the cosmological constant, $w=-1$.  Upper left: The
  baseline fit, where the SNLS3 contours include statistical and all
  identified systematic uncertainties. Upper right: The filled SNLS3
  contours include statistical uncertainties only; the dotted open
  contours refer to the baseline fit with all systematics included.
  Lower left: The filled SNLS3 contours exclude the SN~Ia systematic
  uncertainties related to calibration.  Lower right: The filled SNLS3
  contours result from fixing $\alpha$ and $\beta$ in the cosmological
  fits. See Table~\ref{tab:cosmofits} and
  Table~\ref{tab:syserrorbudget} for numerical
  data.\label{fig:omw_syseffect}}
\end{figure}

The confidence contours for \omatter\ and $w$ in a flat universe can
be found in Fig.~\ref{fig:omw_syseffect} (upper left panel) for fits
considering all systematic and statistical uncertainties.
Fig.~\ref{fig:omw_syseffect} also shows the
statistical-uncertainty-only cosmological fits in the upper right
panel. The best-fitting cosmological parameters and the nuisance
parameters $\alpha$, $\beta$, ${\mathcal M_{B}^1}$ and ${\mathcal
  M_{B}^2}$, for convenience converted to $M_B$ assuming
$H_0=70$\,km\,s$^{-1}$\,Mpc$^{-1}$ (in the grid marginalization
approach, $H_0$ is not fit for as it is perfectly degenerate with
$M_B$), are in Table~\ref{tab:cosmofits_wminus1} (for non-flat, $w=-1$
fits) and Table~\ref{tab:cosmofits} (for flat, constant $w$ fits).  We
also list the parameters obtained with the $\chi^2$ minimization
approach for comparison. All the fits, with and without the inclusion
of systematic errors, are consistent with a $w=-1$ universe: we find
$w=-1.043^{+0.054}_{-0.055}$ (stat), and $w=-1.068^{+0.080}_{-0.082}$
(stat+sys). For comparison, with no external constraints (i.e., SNLS3
only) the equivalent values are $w=-0.90^{+0.16}_{-0.20}$ (stat) and
$w=-0.91^{+0.17}_{-0.24}$ (stat+sys) \citepalias{2011ApJS..192....1C}.

The lower right panel of Fig.~\ref{fig:omw_syseffect} shows the
importance of allowing the nuisance parameters $\alpha$ and $\beta$ to
vary in the fits, rather than holding them fixed at their best-fit
values. This leads to not only smaller contours and hence
underestimated parameter uncertainties, but also a significant bias in
the best-fit parameters (Table~\ref{tab:syserrorbudget}). Holding
$\alpha$ and $\beta$ fixed gives $w=-1.117^{+0.081}_{-0.082}$, a
$\sim0.6\sigma$ shift in the value of $w$ compared to the correct fit.

\begin{figure}
\centering
\plottwo{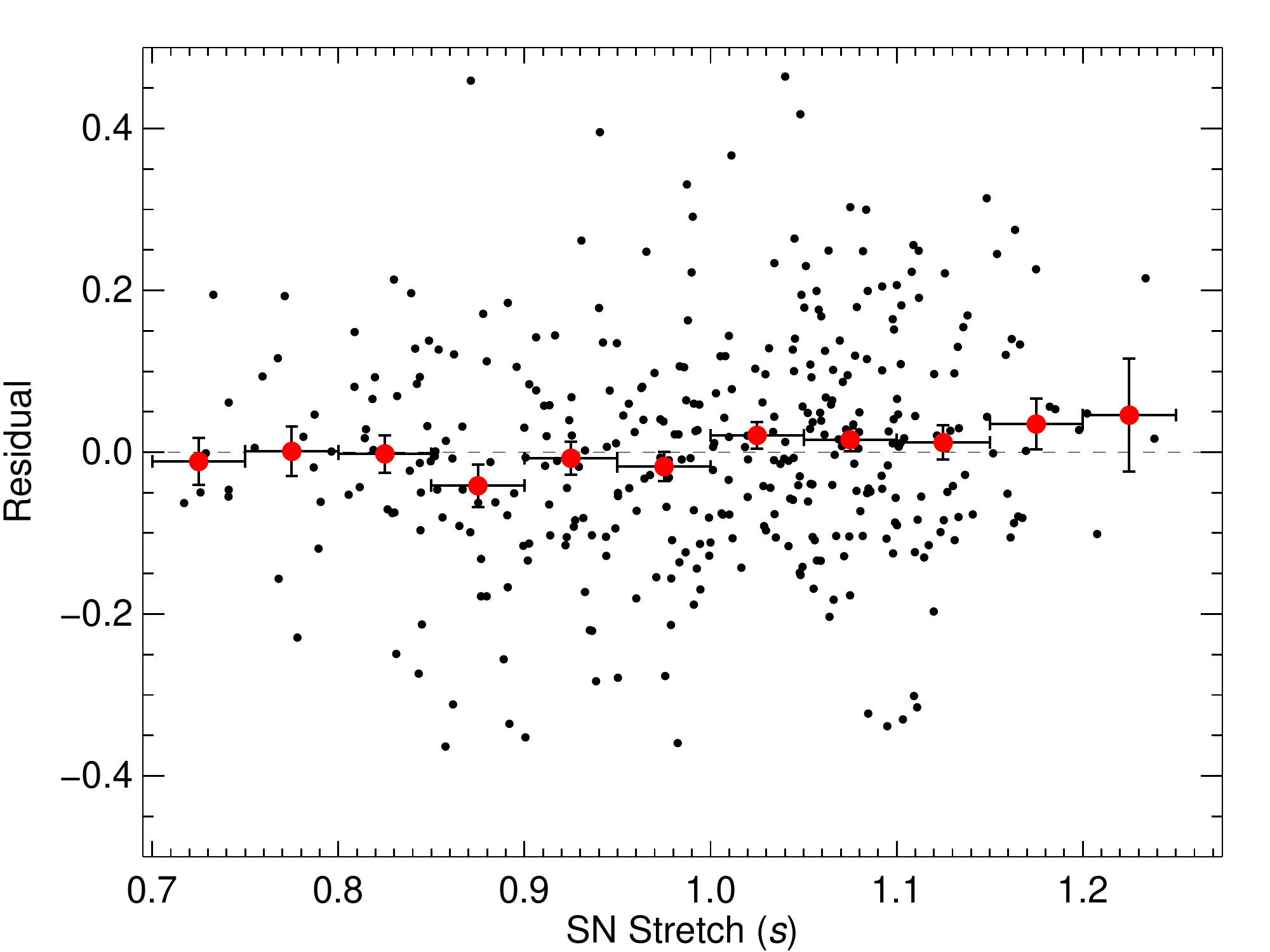}{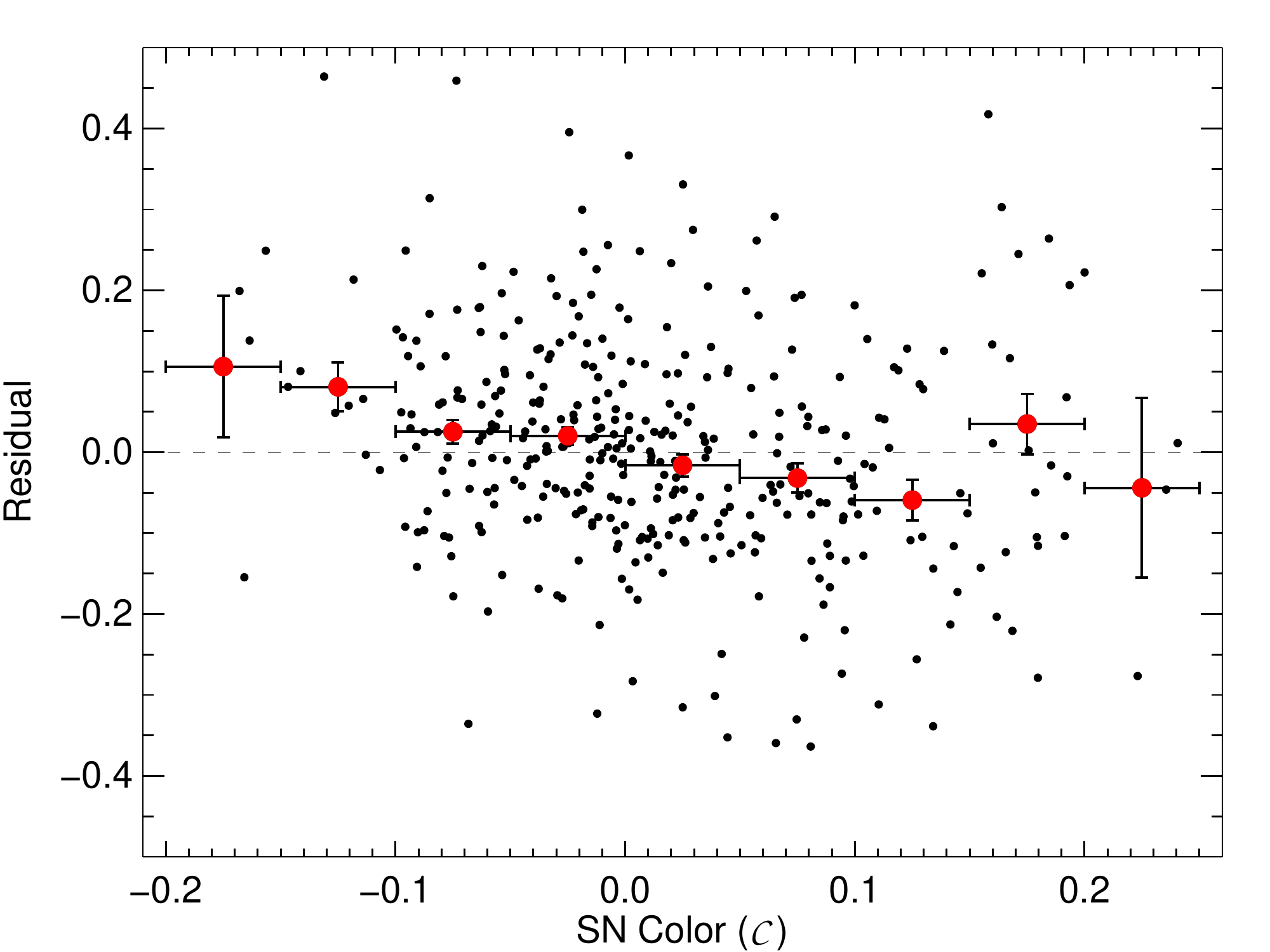}
\caption{Residuals (in magnitudes) from the best-fitting flat
  cosmology as a function of stretch (left) and color (right).
  Residuals are defined as $m_B-\mbmodel$, i.e., negative residuals
  indicate brighter SNe (after application of stretch-- and
  color--luminosity relations). Red points show the mean residuals in
  bins of stretch and color.  The dashed line indicates a zero
  residual.\label{fig:scresid}}
\end{figure}

The residuals from the best-fitting cosmology as a function of stretch
and color can be found in Fig.~\ref{fig:scresid}. No significant
remaining trends between stretch and Hubble residual are apparent, but
there is some evidence for a small trend between SN~Ia color and
luminosity at $\col<0.15$ (indicating that these SNe prefer a smaller
$\beta$, or a shallower slope, than the global value). We examine
this, and related issues, in more detail in
$\S$~\ref{sec:cosm-sub-sampl}.

Covariances between the nuisance parameters are small, with $|r|<0.15$
for most combinations of $\alpha$, $\beta$ and ${\mathcal M_{B}}$. The
exception is between ${\mathcal M_{B}^1}$ and ${\mathcal M_{B}^2}$,
where the correlation is (as expected) larger ($r\sim0.6$). Note that
this positive covariance enhances the significance of the difference
between ${\mathcal M_{B}^1}$ and ${\mathcal M_{B}^2}$ beyond the
simple statistical uncertainties listed in the tables.

\subsection{Systematic error budget}
\label{sec:syst-error-budg}

The \omatter--$w$ flat universe fits (Table~\ref{tab:cosmofits})
represent a 5.2\% statistical measurement of $w$ and a 7.6\% measure
with systematics (i.e., $\simeq5.5$\% with systematics only). The
total systematic uncertainty is therefore comparable to, but slightly
larger than, the statistical uncertainty.  The full systematic
uncertainty error budget can be found in
Table~\ref{tab:syserrorbudget}.  Systematic uncertainties generate
about a $\sim$70\% increase in the size of the area of the
SNLS3+BAO+WMAP7 \omatter--$w$ 68.3\% confidence contour relative to a
fit considering statistical errors only (compared to an 85\% increase
in the SN-only contour; see \citetalias{2011ApJS..192....1C}).

The dominant systematic uncertainty is calibration \citepalias[as
in][]{2011ApJS..192....1C}, and in particular how well-known the
colors and SED of the flux standard (BD 17$^{\circ}$~4708) are -- each
of these two terms provides about a 20\% increase in the contour area
size over the statistical-only fit. The SNLS instrumental zeropoints
and filter responses are also a large effect, generating a $\sim$15\%
increase in each case. In part, this is because the SNLS data are
calibrated to the \citet{1992AJ....104..340L} system (for comparison
to the low-redshift literature SNe), for which the color terms from
the SNLS filters are large \citep{2009A&A...506..999R}. This situation
should improve in the near future as new low redshift SN~Ia samples
observed in a similar filter system to SNLS become available,
dramatically reducing these calibration uncertainties.

By contrast, systematics caused by potential evolution in SN~Ia
properties (the parameters $\alpha$ and $\beta$) are considerably
smaller. As discussed in \citetalias{2011ApJS..192....1C}, we find no
evidence that $\alpha$ varies with redshift, and only marginal
evidence for redshift variation in $\beta$: explicitly
$d\alpha/dz=0.021\pm0.07$ and $d\beta/dz=0.588\pm0.40$
\citepalias{2011ApJS..192....1C}. Is is unlikely that this $\beta$
evolution is real \citepalias{2010A&A...523A...7G}; however, we
conservatively adopt $d\alpha/dz=0.07$ (the uncertainty in the slope)
and $d\beta/dz=1.0$ in our systematics analysis. Even this amount of
redshift evolution in $\alpha$ and $\beta$ contributes an almost
negligible effect (Table~\ref{tab:syserrorbudget}). The largest
identified systematic uncertainty related to the astrophysics of SNe
Ia is the implementation of the host-galaxy dependent term in
eqn.~(\ref{eq:mtheory}).

The lower left panel of Fig.~\ref{fig:omw_syseffect} shows the
\omatter--$w$ contours with all systematics included, \textit{except}
those related to calibration. These ``no-calibration-systematics''
contours are very similar to the statistical-only contours (only a
factor 1.07 larger), with $w=-1.048\pm^{+0.057}_{-0.058}$. This
represents a total error in $w$ of $\simeq5.5$\%, and a systematic
contribution of $\simeq1.8$\%, significantly smaller than when the
calibration systematics are included. With our current knowledge and
fitting techniques for SNe Ia, this represents the systematic floor
given a negligible photometric calibration uncertainty.

\subsection{Cosmological results}
\label{sec:cosmo-results}

We now present our main cosmological results. We consider various
combinations of the SNLS3, WMAP7, SDSS DR7 LRGs, and $H_0$ datasets:
WMAP7+SNLS3+DR7 is the most similar to the constraints used in the
grid marginalization approach (Table~\ref{tab:cosmofits}), but still
differs as it uses the full matter power spectrum of LRGs rather than
the BAO constraint, and the full WMAP7 power spectrum rather than the
shift parameters. The best-fitting value of $w$
(Table~\ref{tab:cosmomcfits_constw}) is therefore slightly different
due to these differing external constraints even though the SN~Ia
constraints are identical, but the percentage error in $w$ is the same
at 7.6\%.

\subsubsection{Constant $w$ fits}
\label{sec:const-w}

\begin{figure*}
\plotone{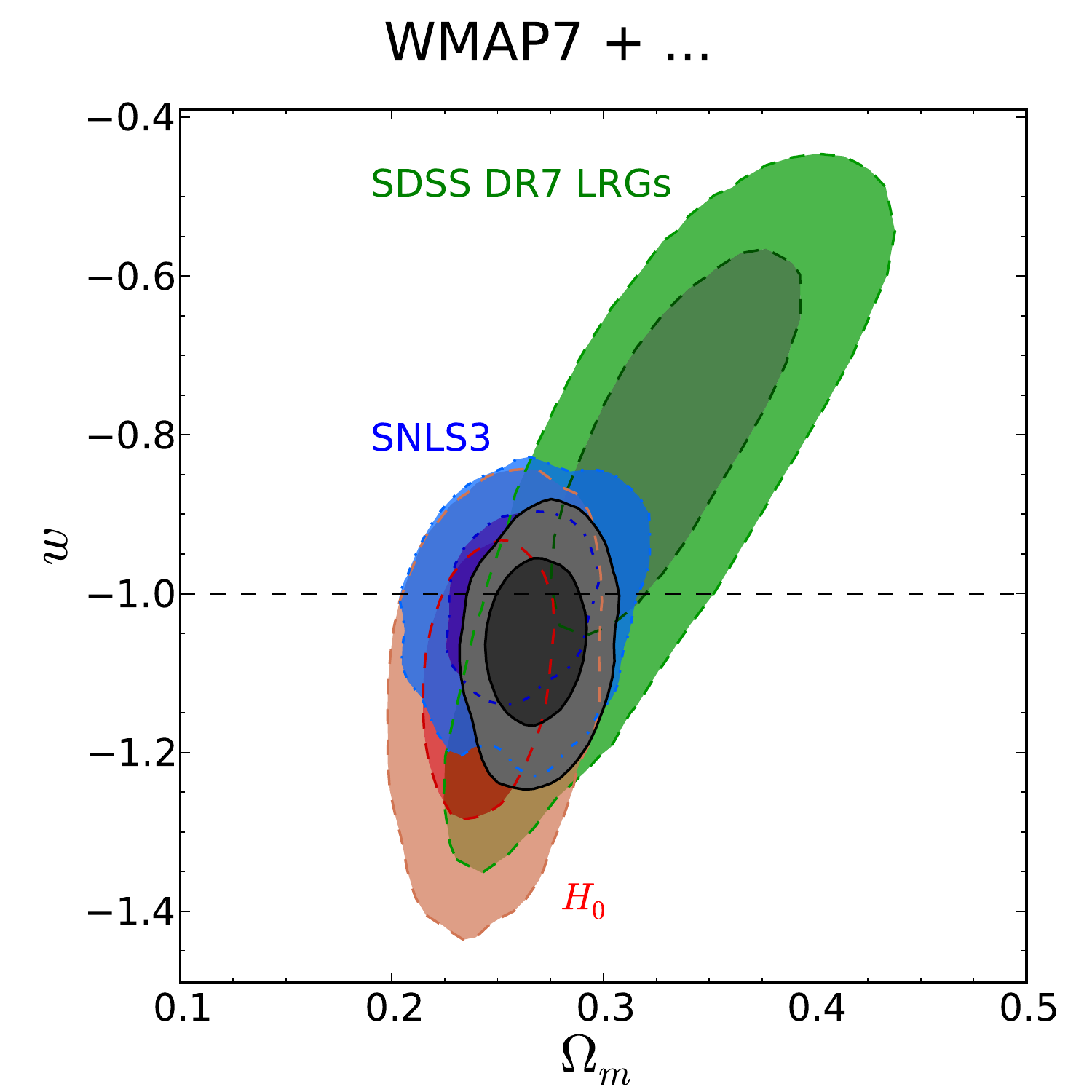}
\caption{Confidence contours in the cosmological parameters \omatter\
  and $w$ assuming a flat universe, produced using the
  \texttt{CosmoMC} program. The SNLS3 contours are in blue, the SDSS
  DR7 LRG contours in green, and the $H_0$ prior in red.  WMAP7
  constraints are included in all contours.  The contours enclose
  68.3\% and 95.4\% of the probability and include all SN systematic
  uncertainties. The dashed line indicates $w=-1$. Numerical results
  are in
  Table~\ref{tab:cosmomcfits_constw}.\label{fig:omw_flat_cosmomc}}
\end{figure*}

\begin{figure*}
\plotone{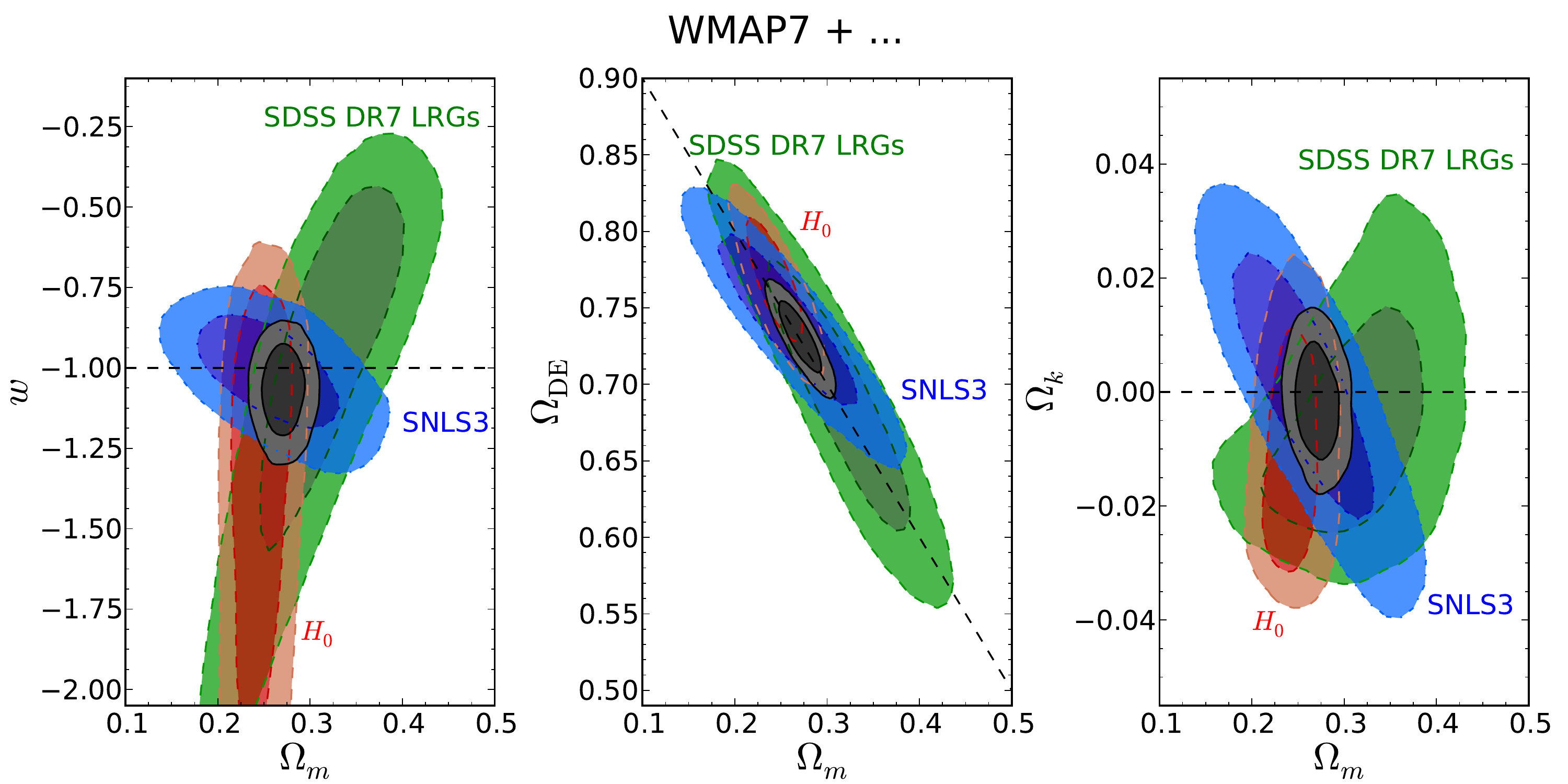}
\caption{Confidence contours in the cosmological parameters \omatter,
  \ode, \ok, and $w$ produced using the \texttt{CosmoMC} program. The
  SNLS3 contours are in blue, the SDSS DR7 LRG contours in green, and
  the $H_0$ prior in red.  WMAP7 constraints are included in all
  contours.  The contours enclose 68.3\% and 95.4\% of the probability
  and include all SN systematic uncertainties. In the left-hand panel,
  the dashed line indicated $w=-1$; in the center and right-hand
  panels the line indicates a flat ($\ok=0$) universe. Numerical
  results are in
  Table~\ref{tab:cosmomcfits_constw}.\label{fig:omw_nonflat_cosmomc}}
\end{figure*}

All the results are consistent with a spatially flat, $w=-1$ universe.
Our results for a flat universe with a constant dark energy equation of state are
\begin{eqnarray}
\nonumber \omatter&=&0.269\pm{0.015}\\
\nonumber w&=&-1.061^{+0.069}_{-0.068},
\end{eqnarray}
and, relaxing the assumption of spatial flatness,
\begin{eqnarray}
\nonumber \omatter&=&0.271\pm{0.015}\\
\nonumber \ok&=&-0.002\pm{0.006}\\
\nonumber w&=&-1.069 ^{+0.091}_{-0.092},
\end{eqnarray}
including external constraints from WMAP7 and SDSS DR7 and a prior on
$H_0$ (all quoted uncertainties in this section include both the SN
statistical and systematic components). The confidence contours are in
Fig.~\ref{fig:omw_flat_cosmomc} and
Fig.~\ref{fig:omw_nonflat_cosmomc}, and the corresponding best-fit
cosmological parameters for various combinations of external
constraints can be found in Table~\ref{tab:cosmomcfits_constw}. 

In Table~\ref{tab:cosmomc_allparams} we give a full list of all the
best-fit parameters from the constant $w$ \texttt{CosmoMC} fits with
the WMAP7, SDSS DR7 and $H_0$ external datasets\footnote{Full
  parameter summaries for all combinations of external datasets can be
  found at \url{https://tspace.library.utoronto.ca/snls}.}. This
includes some parameters which the SN Ia data do not directly
constrain. $\Omega_b$ and $\Omega_c$ are the fractional energy
densities of baryons and dark matter, $\tau$ is the reionization
optical depth, $n_s$ is the scalar spectral index, $A_{05}$ the
amplitude of curvature perturbations at $k=0.05$Mpc$^{-1}$, and
$\sigma_8$ the normalization of the matter power spectrum at
$8h^{-1}$\,Mpc.

Of particular note is the high importance of the SN~Ia dataset in
placing meaningful constraints on $w$. Assuming a flat universe,
WMAP7+DR7 alone only measure $w$ to $\sim$20\%, and adding the $H_0$
prior (i.e., WMAP7+DR7+$H_0$) only decreases this uncertainty to
$\sim$11\%.  Including the SNLS3 dataset with WMAP7+DR7, by contrast,
reduces the uncertainty to 7.7\%.  WMAP7+SNLS3 together also provide a
7.7\% measurement. With all external constraints, including the $H_0$
prior, the measurement of $w$ is 6.5\%, comparable to WMAP7+SNLS3
alone. Note that the DR7 constraint has almost no effect on the
uncertainty in the measurement of $w$ -- WMAP7+$H_0$+SNLS3 has
essentially the same uncertainty as WMAP7+DR7+$H_0$+SNLS3.

The situation is slightly different when making no assumption about
spatial flatness, but the basic result of the high importance of the
SN~Ia data remains. In this case, the WMAP7+DR7+SNLS3 data provide an
8.8\% measurement, compared to 11.3\% without DR7 (the DR7 data make
important contributions towards constraining \omatter).
WMAP7+DR7+$H_0$ alone can only make a $\sim20$\% measurement; adding
SNLS3 improves this dramatically to $\sim8.5$\%.

\subsubsection{Variable $w$ fits}
\label{sec:varying-w}

\begin{figure*}
\plotone{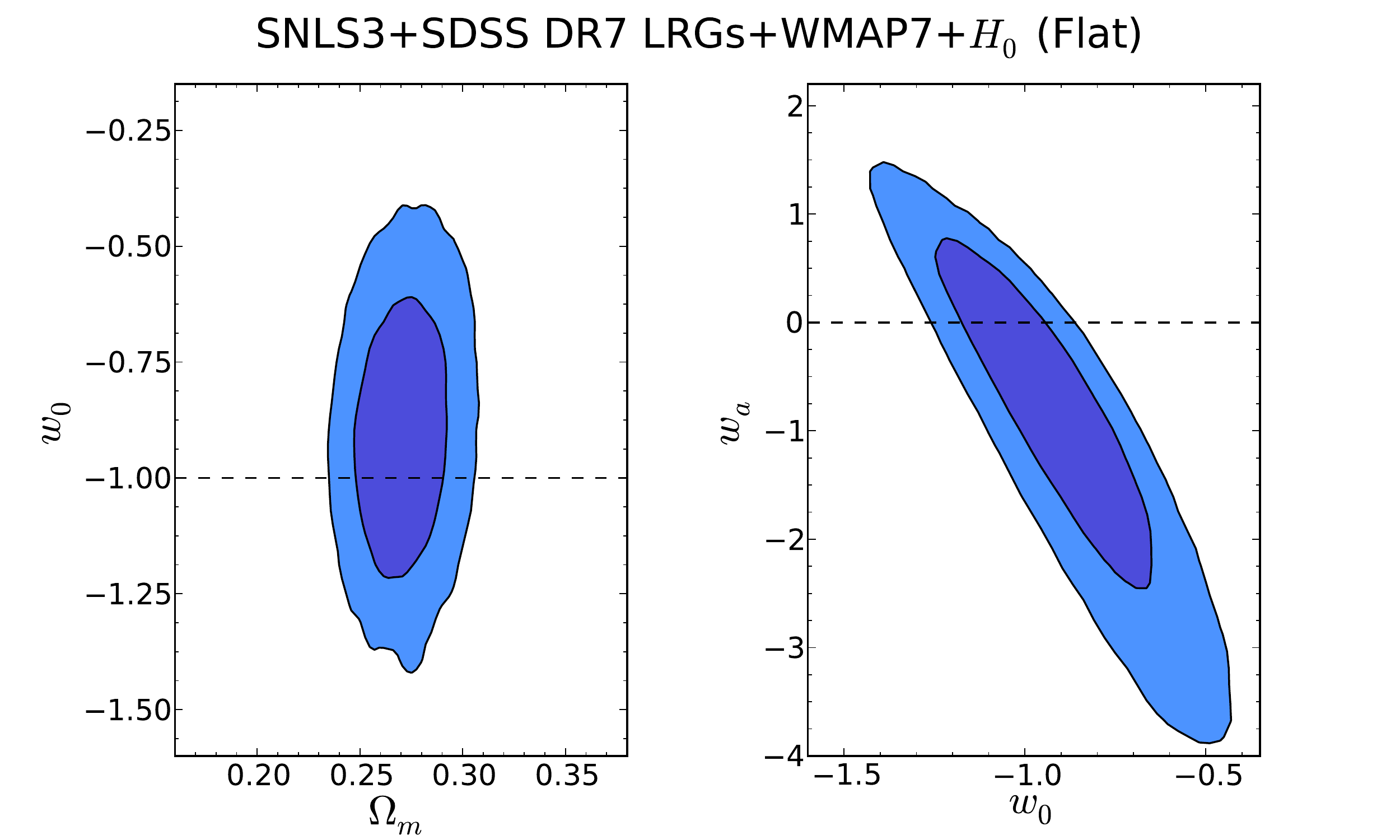}
\caption{Combined confidence contours in \omatter, \wo, and \wa\ using
  SNLS3, WMAP7, SDSS DR7 LRGs, and a prior on $H_0$. A flat universe
  is assumed, and we enforce a prior of $\wo+\wa\leq0$ -- any apparent
  discrepancy with this prior is a result of smoothing the CosmoMC
  output. The horizontal dashed lines indicate a cosmological constant
  ($\wo=-1$; left) and a non-varying $w$ ($\wa=0$; right).  All SN Ia
  systematic uncertainties are included. Numerical results are in
  Table~\ref{tab:cosmomcfits_varw}.\label{fig:omwa_flat_cosmomc}}
\end{figure*}

\begin{figure*}
\plotone{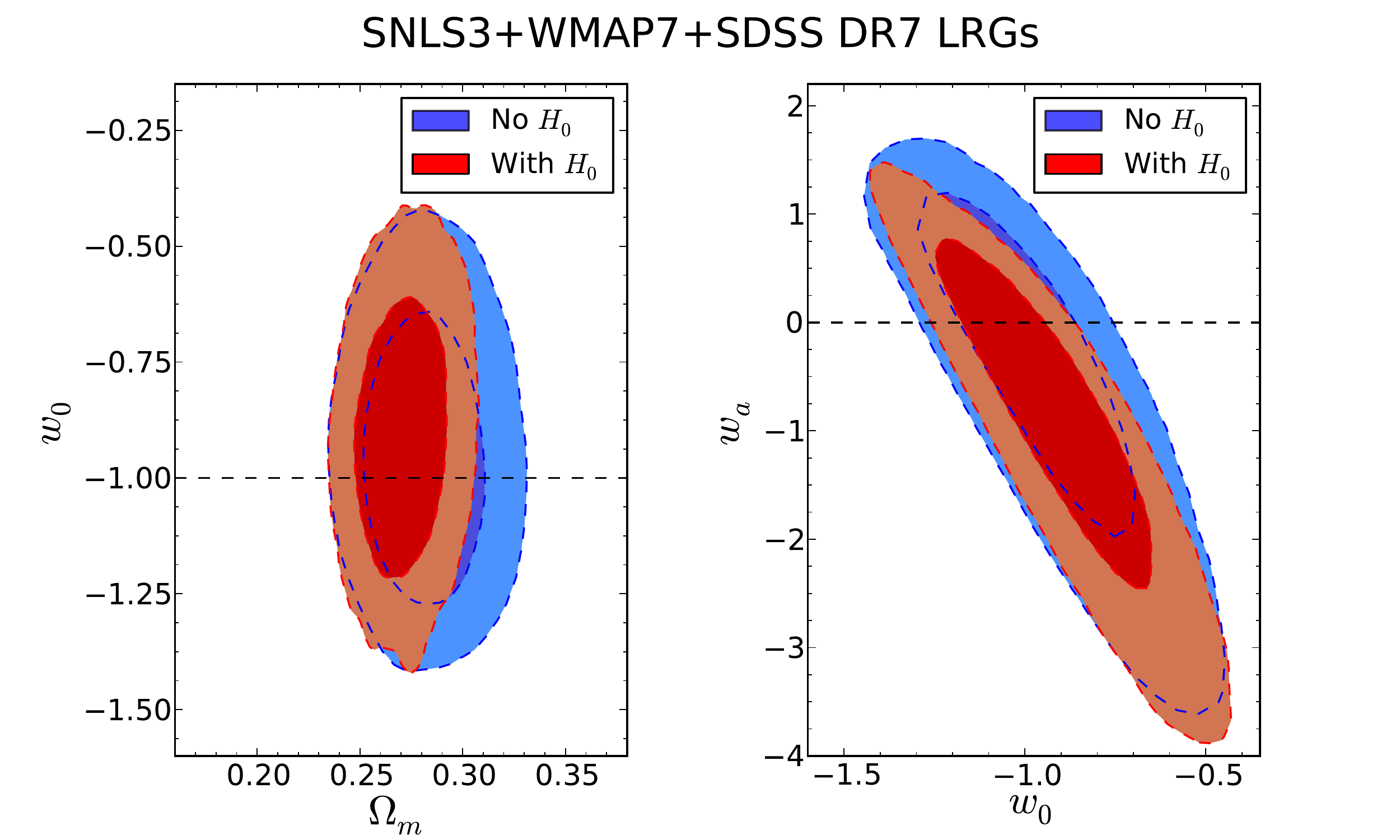}
\caption{The effect of the $H_0$ prior on the \omatter, \wo, and \wa\
  contours in a flat universe. The red contours show the fits with the
  $H_0$ prior, and the blue contours
  without.\label{fig:omwa_noH0_cosmomc}}
\end{figure*}

The final set of fits allow the equation of state parameter $w$ to
vary simply as a function of the scale factor, $a$, as
$w(a)=\wo+\wa(1-a)$ , with a cosmological constant equivalent to
$\wo=-1$, $\wa=0$. We use a hard prior of $\wo+\wa\leq0$, from the
constraint of matter domination in the early universe. The confidence
contours are in Fig.~\ref{fig:omwa_flat_cosmomc} assuming a flat
universe. The best-fit parameters are listed in
Table~\ref{tab:cosmomcfits_varw}.  Again, we find no evidence of
deviations from the cosmological constant. Assuming a flat universe we
find
\begin{eqnarray}
\nonumber \omatter&=&0.271^{+0.015}_{-0.015}\\
\nonumber \wo&=&-0.905^{+0.196}_{-0.196}\\
\nonumber \wa&=&-0.984^{+1.094}_{-1.097}.
\end{eqnarray}
The $H_0$ prior has only a small effect on the $w(a)$ fits
(Fig.~\ref{fig:omwa_noH0_cosmomc}). As a comparison, with no $H_0$
prior, we find $\wo=-0.949^{+0.198}_{-0.201}$,
$\wa=-0.535^{+1.109}_{-1.111}$. The SN~Ia data are critical for a
constraining measurement -- without the SN data, fits for \wo\ and
\wa\ did not converge.

\subsection{Comparison to other results}
\label{sec:comparison}

We compare our results to previous constraints on dark energy using SN
Ia data. \citet{2011ApJS..192...18K}, with a combination of WMAP7,
BAO, the \citet{2009ApJ...699..539R} $H_0$ measurement, plus the
\citet{2009ApJ...700.1097H} SN~Ia dataset, found $\wo=-0.93\pm0.13$
and $\wa=-0.41^{+0.72}_{-0.71}$. Although this may appear to be better
than our constraints, it did not include a proper handling of SN~Ia
systematics due to the lack of a consistent, published dataset at that
time, and the uncertainties will therefore be under-estimated. Using a
slightly larger SN~Ia dataset (``Union2'') the same WMAP7/BAO
constraints, and the \citet{2009ApJ...699..539R} $H_0$ measurement,
\citet{2010ApJ...716..712A} find $\omatter=0.274^{+0.016}_{-0.015}$,
$\ok=-0.002\pm0.007$, and $w=-1.052^{+0.092}_{-0.096}$, comparable to
our results (\S~\ref{sec:cosmo-results}). However, those authors left
$\alpha$ and $\beta$ fixed when computing systematic uncertainties,
which may underestimate the size of the final uncertainties
\citepalias[][see also Fig.~\ref{fig:omw_syseffect} and
Table~\ref{tab:syserrorbudget}]{2011ApJS..192....1C}, and likely
under-estimated the magnitude of the photometric calibration
uncertainties (see \citetalias{2011ApJS..192....1C} for discussion).

We also compare our results with those predicted by the Dark Energy
Task Force \citep[DETF;][]{2006astro.ph..9591A} for an experiment of
this type by calculating the figure of merit (FoM) for our combination
of datasets.  The exact definition of the FoM has some ambiguity: in
the DETF report it is defined as proportional to the reciprocal of the
area of the error ellipse in the \wo--\wa\ plane that encloses 95\% of
the total probability, but the constant of proportionality is never
stated. The values given in \citet{2006astro.ph..9591A} are based on
performing a transform of variables from \wo\ to $w_p$, the value of
$w$ at the so-called pivot redshift $z_p$, where $w_p$ and \wa\ are
uncorrelated.  The FoM is then simply taken to be
$1/(\sigma_{w_p}\sigma_{\wa})$.  This prescription is the most
commonly used in the literature \citep[e.g.,][]{2011arXiv1101.1529E},
although some authors have been more literal in taking the area of the
ellipse.  FoM calculations should not assume a flat universe; however
current data are poorly constraining without this constraint, so we
follow the practice in the literature and assume flatness in our FoM
numbers.

For our WMAP7+DR7+SNLS3+$H_0$ fit in a flat universe, we find
$z_p\simeq0.19$ and $w_p=-1.063\pm0.082$.  Combined with our
measurement of $w_a$, we find a FoM of 11.1 (see also
Table~\ref{tab:cosmomcfits_varw}).  Excluding the SHOES $H_0$ prior
gives a FoM of 10.6. Directly taking the reciprocal of the area of the
95\% confidence intervals give 0.56 and 0.44 respectively.

In DETF terminology, the SNLS3 sample represents a stage II SN
experiment, and the final combination of this and other stage II
experiments is predicted to give a FoM of $\simeq50$. However, these
figures are difficult to compare with our results. The DETF
calculations assume a far larger SN~Ia sample of 1200 events
(including, significantly, 500 at low-redshift), together with a
(superior) CMB prior from the Planck satellite
\citep{2011arXiv1101.2022P} rather than WMAP, and also include cluster
and weak lensing experiments. They do not include BAO or $H_0$
information, although the latter only has a small effect on the FoM.

\section{Supernova sub-samples}
\label{sec:cosm-sub-sampl}

A complementary approach to checking and analyzing systematics, in
particular those that are not susceptible to an analytical approach,
is to break the SN~Ia sample into sub-samples which probe either
different experimental systematics, or different regions of parameter
space of the SN~Ia population. For example, the stretch--luminosity
and color--luminosity relations are assumed to be linear, universal
and invariant with (e.g.) SN properties, and the size of the SNLS3
sample allows us to test these assumptions in detail. Furthermore, the
cosmological results should be robust to any segregation of the data
if the systematics are handled correctly.

For simplicity, we use the $\chi^2$ minimization approach in this
section. We first split the SNLS SNe according to location on the sky
(i.e., one of the four CFHT-LS deep fields). We then test the
robustness of the nuisance parameters by splitting the sample by SN
and their host galaxy parameters.

\subsection{Segregation by SNLS field}
\label{sec:segr-snls-field}

We first test for any variations in $w$ as a function of the field in
which the SNLS SN occurred. SNLS observes in four fields distributed
in right ascension \citep[see][for the field
coordinates]{2006AJ....131..960S}, so we can test for a combination of
photometric calibration systematics, as well as a physically varying
$w$ in different directions, by comparing the cosmological parameters
we derive from each field. To do this, we adjust our $\chi^2$
minimization approach to fit for global nuisance parameters ($\alpha$,
$\beta$, and \scriptm) for all SNe, but with a different $w$ and
\omatter\ in each of the four different SNLS fields. The average $w$
and \omatter\ for the four fields is applied to the external SNe and
for combination with WMAP7 and the BAO constraints.

This approach effectively adds six new terms to the fit. The results
can be found in Table~\ref{tab:fieldvar}. The $\chi^2$ only drops from
418.1 (for 466 DOF) to 414.9 (for 460 DOF), indicating that the fields
are consistent. Comparing the individual field \omatter\ and $w$ values
to the average values of the four fields, and allowing for the
covariances between the individual values, gives a $\chi^2$ of 3.38
for 6 DOF, consistent with the $\chi^2$ distribution and indicating no
significant variability among the different fields.

\subsection{Segregation by SN properties}
\label{sec:segr-snprops}

The SN~Ia light curve shape is well-known to vary systematically as a
function of the SN environment. Fainter SNe~Ia with faster light
curves are preferentially located in older stellar populations, while
the brighter examples with broad light curves tend to explode in
late-type spiral or star-forming systems
\citep[e.g.,][]{1995AJ....109....1H,2000AJ....120.1479H,2006ApJ...648..868S}.
When coupled with the evidence for both a young and old component to
the SN~Ia progenitor population
\citep{2005A&A...433..807M,2006MNRAS.370..773M,2006ApJ...648..868S,2010AJ....140..804B},
or at least a wide-range in the SN delay-times
\citep{2008ApJ...683L..25P,2008PASJ...60.1327T}, a natural prediction
is a subtle change in the mix of SN light curve shapes with redshift
\citep{2007ApJ...667L..37H}.

If the stretch--luminosity relation is universal across SN stretch and
progenitor age, this predicted drift will not impact the determination
of the cosmological parameters -- a low-stretch SN should correct
equally well as a high-stretch SN. To test this, we split our SN
sample into two groups ($s$$<$1 and $s$$\geq$1) and perform
independent cosmological fits to each sub-sample. This obviously
restricts the lever-arm in stretch and so the $\alpha$ coefficient is
less well-determined; nonetheless this is a useful test of the utility
of SNe~Ia across different environments.

When considering sub-samples of SNe classified by SN properties (for
example stretch or color), the Malmquist corrections that are applied
globally to the sample will not be appropriate. For example, low
stretch SNe Ia are intrinsically fainter and will suffer from a larger
selection effect at high-redshift than high stretch events.  Rather
than apply different Malmquist corrections for these different
sub-samples \citep[which could in principle be derived from
simulations such as those in][]{2010AJ....140..518P}, we instead
restrict the SNe to the redshift ranges over which selection effects
are reduced.  For the SNLS sample, we restrict to $z<0.75$, and for
the SDSS sample, we restrict to $z<0.3$. These data also tend to be
better observed, being brighter, with smaller error bars on the SN
parameters. We also discard the small HST sample. We perform both
statistical and statistical+systematic uncertainty fits -- the
comparison is useful as we are interested in the differences between
SN~Ia sub-samples, and many of systematics affect sub-samples in a
similar way.

\begin{figure}
\plotone{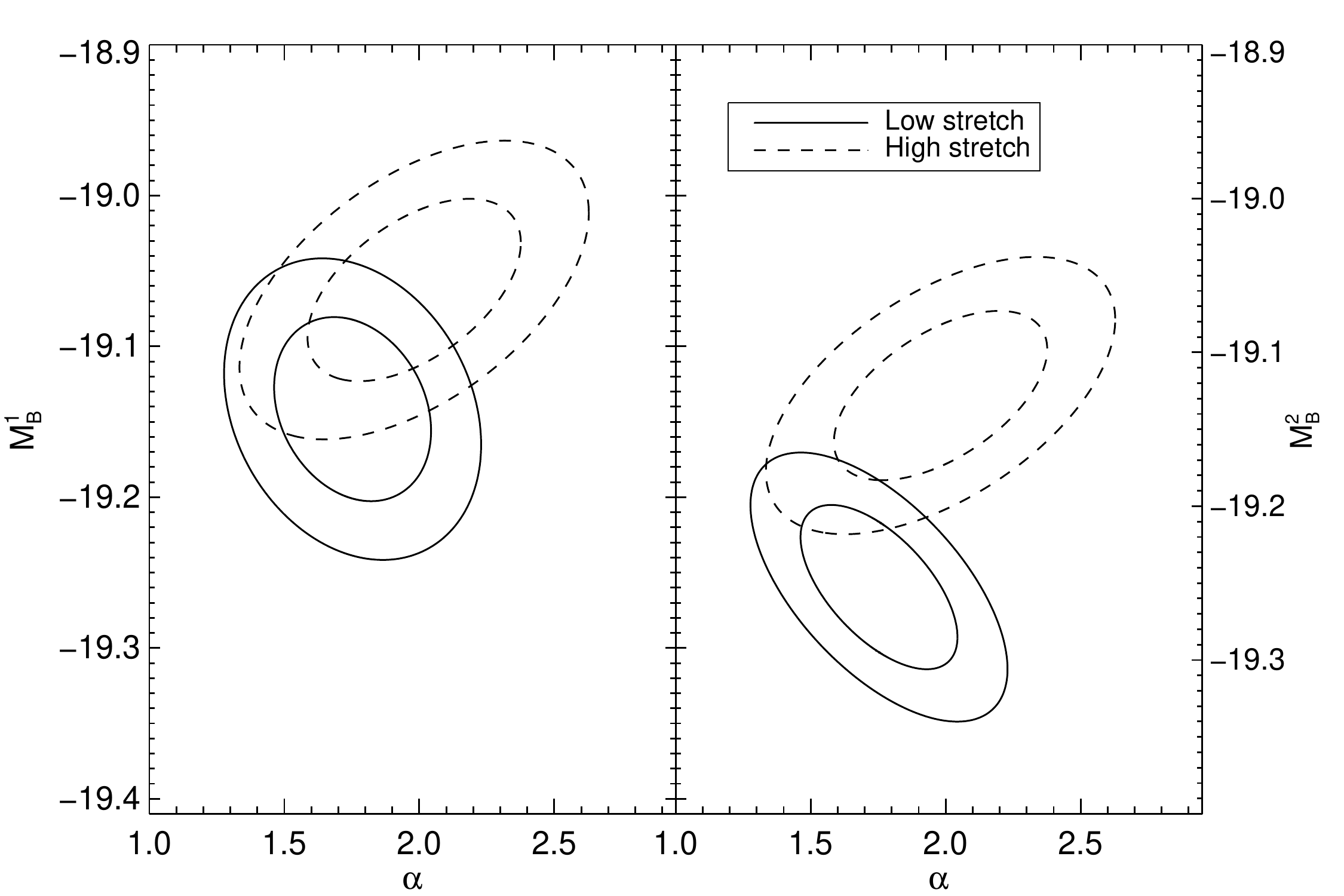}
\caption{Joint confidence contours between the nuisance parameters
  $\alpha$, $M_B^1$ (left), and $M_B^2$ (right) for low-stretch (solid
  line) and high-stretch (dashed line) SNe Ia, using the restricted SN
  Ia sample described in $\S$~\ref{sec:segr-snprops}. The contours
  enclose 68.3\% and 95.4\% of the probability, and the fits include
  all systematic uncertainties. Only mild tensions
  exist.\label{fig:svar_a_m1}}
\end{figure}

The results are given in Table~\ref{tab:cosmofitssubsample} and
Fig.~\ref{fig:svar_a_m1}, the latter generated from the covariance
matrices of the fits.. The derived cosmological parameters are
consistent between low and high-$s$ SNe Ia, although the nuisance
parameters show some differences: the $\alpha$ values are consistent
between low-$s$ and high-$s$, but much larger than the full sample.
The $M_B$ values show large differences between low-$s$ and high-$s$.
However, there are significant (and expected) covariances between
$\alpha$ and $M_B$ in these fits: for the low-$s$ group, $M_B^1$
increases as $\alpha$ decreases, while for the high-$s$ group $M_B^1$
decreases for decreasing $\alpha$ (and a similar trend is seen for
$M_B^2$). As $M_B$ is defined at $s=1$, which the SNe in neither group
sample, the two parameters become quite inter-dependent. When
considering the joint confidence contours between $\alpha$, $M_B^1$,
and $M_B^2$ (Fig.~\ref{fig:svar_a_m1}) only mild tensions are seen. A
slightly larger tension is seen in the value of $\beta$ for the two
sub-groups: the values $2.90\pm0.16$ and $3.25\pm0.14$ differ at
$\sim2.2\sigma$ (Table~\ref{tab:cosmofitssubsample}).

In a similar vein, we also split the sample by SN color at $\col=0$.
Unlike stretch, SN color (examined independently of SN luminosity)
shows no clear evidence for variation with environment. No robust
trends have been found despite rigorous examinations
\citep[e.g.,][]{2010MNRAS.406..782S,2010ApJ...722..566L}, and we also
find no significant trends of the nuisance parameters varying as a
function of SN color.

\subsection{Segregation by host galaxy characteristics}
\label{sec:segr-host-galaxy}

We also segregate the SNLS3 sample according to the environment in
which the SN exploded. This is a fundamentally different test,
dividing the sample according to the environment of the SN rather than
its direct properties (although significant correlations exist
between, for example, SN stretch and host galaxy \mstellar\ or
star-formation rate). For these tests we use \mstellar\ to segregate
the sample, as used in the cosmological fits, following
\citet{2010MNRAS.406..782S}.  As we already use a different $M_B$ for
SNe Ia in low and high \mstellar\ galaxies in our standard fits, we
simply adapt this approach to additionally fit for a different
$\alpha$ and $\beta$ in the two host classes. This has the advantage
that we are no longer comparing the results of different fits as we
fit for all the nuisance parameters simultaneously, and we can use the
full systematics covariance matrix, as well as the full SN~Ia sample.
(We also provide the fit results when we physically divide the sample
into two, for comparison with $\S$~\ref{sec:segr-snprops}, in
Table~\ref{tab:cosmofitssubsample}.) Note that in these fits we do not
include the host galaxy systematic term, as we are trying to examine
the effect of any host galaxy dependence.

\begin{figure}
  \plotone{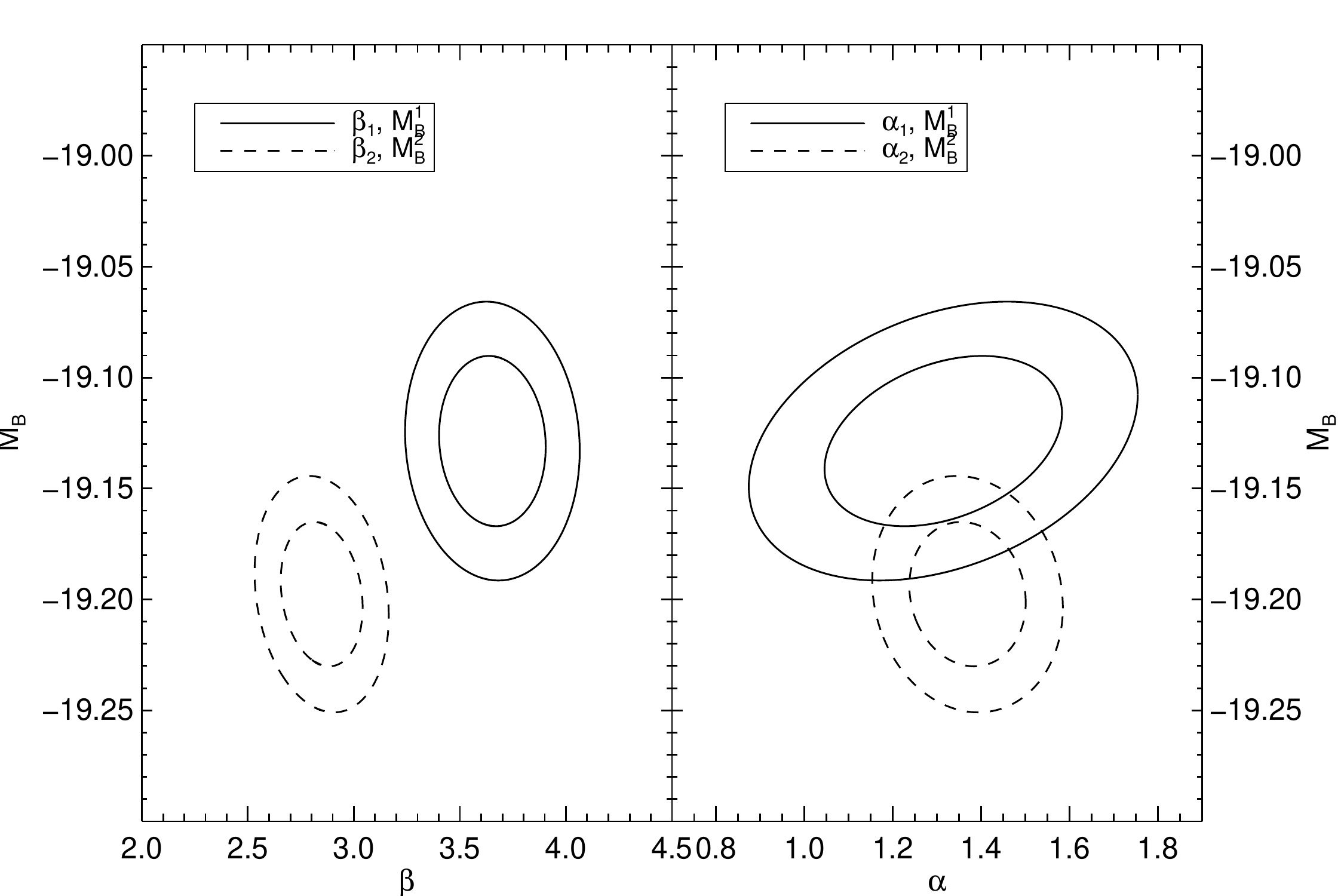}
  \caption{Joint confidence contours in the nuisance parameters
    $\beta$ and $M_B$ (left) and $\alpha$ and $M_B$ (right), allowing
    all the nuisance parameters to vary according to host galaxy
    stellar mass. $\alpha_1$/$\beta_1$/$M_B^1$ refer to SNe Ia in
    hosts with $\mstellar\le10^{10}\mathrm{M}_{\odot}$, and
    $\alpha_2$/$\beta_2$/$M_B^2$ to SNe Ia in hosts with
    $\mstellar>10^{10}\mathrm{M}_{\odot}$. The full SNLS3 sample is
    used, and all systematic uncertainties are included. A significant
    variation in $\beta$ with host \mstellar\ is
    observed.\label{fig:multinuisance}}
\end{figure}

The ``multi nuisance parameter'' fit results are shown in
Table~\ref{tab:multi_nuisance}, which gives the values of the nuisance
parameters themselves, and Table~\ref{tab:multi_cosmo}, which gives
the effect on $w$ and the $\chi^2$. The joint confidence contours for
some of the nuisance parameters are shown in
Fig.~\ref{fig:multinuisance}. For completeness, we also give results
when only one $M_B$ is used. 

The data do not support the addition of a different $\alpha$ parameter
in low and high mass galaxies -- when this is added to the fits, the
$\alpha$ values are generally consistent and the quality of the fit,
as indicated by the $\chi^2$, is unchanged. This is true even when
only one $M_B$ is used in the fits, and suggests that $\alpha$ is
fairly insensitive to the details of the environment and
characteristics of the SN~Ia progenitor stellar population.

However, there is evidence for different $M_B$ (as already fit for)
and different $\beta$. The value of $\beta$ is $\sim3.7$ in low mass
galaxies, versus $\sim2.8$ in high-mass galaxies, regardless of
whether two $M_B$ are used.  A similar trend is seen when physically
dividing the sample into two and performing separate independent fits
as in the previous section (Table~\ref{tab:cosmofitssubsample}), and
is consistent with the $\beta$ difference seen between low and
high-stretch SNe in $\S~\ref{sec:segr-snprops}$, as low-stretch SNe
are preferentially found in massive host galaxies.  Generally, the two
$\beta$ values show only a very small positive covariance, and differ
at the $\sim4.3\sigma$ level. There is also a substantial reduction in
the $\chi^2$ of the fit when including two $\beta$ terms. For example,
fitting for two $\beta$s and two $M_B$ reduces the $\chi^2$ to 405.4
from the 423.1 obtained if only one $\beta$ is used (for 465 and 466
DOF). An F-test indicates that this additional term is required at
$\simeq4.5\sigma$.

The variation of $\beta$ with host properties has been observed at
lower significance by \citet{2010MNRAS.406..782S} in the SNLS sample,
and by \citet{2010ApJ...722..566L} in the SDSS SN~Ia sample (a larger
SDSS SN~Ia sample than the one used in this paper).
\citet{2010ApJ...722..566L} find $\beta\sim2.5$ in passive host
galaxies, and $\beta\sim3.1$ in star-forming host galaxies. Using
their full sample, the significance is $\sim3.5\sigma$, although this
drops to $<2\sigma$ when considering only cosmologically useful events
with normal stretches and colors using similar cuts to those used in
this paper.

\subsection{Discussion}
\label{sec:discussion}

The most significant result from our analysis of the SN sub-samples is
the additional variation of $\beta$, as well as $M_B$, between low-
and high-\mstellar\ host galaxies. This effect appears real in our
data and so should be accounted for appropriately in our cosmological
results. We therefore examine the systematic effect of \textit{not}
including this term, and compare to our existing systematic
uncertainty error budget.

Compared to a statistical uncertainty only fit, the addition of two
$\alpha$s and two $\beta$s, gives a $\Delta\omatter=0.00$ and $\Delta
w=0.005$ (Table~\ref{tab:multi_cosmo}). The mean statistical-only
errors on \omatter\ and $w$ are 0.0148 and 0.0545
(Table~\ref{tab:syserrorbudget}). Adding in quadrature the shifts
measured when including the two $\alpha$s and $\beta$s increases the
$w$ uncertainty to 0.0547. This total uncertainty on $w$ is smaller
than the uncertainty obtained when including host galaxy systematic
term listed in (0.0559; Table~\ref{tab:syserrorbudget}), i.e.  the
effect on $w$ of introducing two $\beta$s is smaller than our current
host galaxy systematic term.

Note this would not be the case if we had neglected all nuisance
parameter variation, i.e. had only used one $M_B$ in our cosmological
fits. In this case, $\Delta\omatter=0.005$ and $\Delta w=0.055$ (the
shift from considering one $M_B$ to considering two $M_B$); in the
case of $w$ this is a shift larger than our statistical uncertainty,
and is comparable to our total systematic uncertainty, becoming the
dominant term in the error budget. Thus while the use of different
$M_B$ is essential for a SN~Ia cosmological analysis, the use of two
$\beta$s \textit{and} two $M_B$s is not. Note that no previous SN Ia
cosmological analysis has performed this host galaxy correction,
indicating that systematic uncertainties will be significantly
under-estimated in these studies.

A similar argument can be made using the fits including systematic
uncertainties. Here, we compare fits that do include the host galaxy
systematic term (unlike the numbers in Table~\ref{tab:multi_cosmo}),
as we wish to examine whether the size of any shift in the
cosmological parameters with the addition of two $\beta$s is accounted
for by our existing systematic uncertainty error budget. In this case,
$\Delta w=0.015$. Our total error in $w$ is 0.0810, compared to 0.0800
excluding the host systematic term (see
Table~\ref{tab:syserrorbudget}). Adding the 0.015 in quadrature to
this 0.0800 gives a $w$ uncertainty of 0.0813, a total uncertainty on
$w$ almost the same as that obtained when using the host systematic
term.  Therefore, our conclusion is that while the two $\beta$ effect
appears real in our data, it is adequately accounted for by our
systematic uncertainty error budget.

Although the variation of $\beta$ with host parameters is not a
concern for this cosmological analysis, it does have implications for
the physical origin of color variation in SNe Ia which may impact
future surveys. A long-standing observation is that the slope of the
relation between $M_B$ and $B-V$ (i.e., $\beta$) is $\ll4.1$
\citep{1998A&A...331..815T,2006A&A...447...31A}, the value expected
based on Milky Way like dust if $\beta$ is interpreted as the ratio of
total-to-selective extinction $R_B$ (where $R_B\equiv R_V+1$, and
$R_V\simeq3.1$ for the Milky Way). The effective $\beta$ for SNe Ia is
likely a conflation of different physical effects, including
extinction by dust \citep[which may vary with host type, e.g.,][and
this paper]{2010ApJ...722..566L,2010MNRAS.406..782S} and intrinsic
variation in SN color that does not correlate with SN light curve
shape \citep[e.g.,][]{2010AJ....139..120F}, and which may depend on
variables such as explosion asymmetry or observational viewing angle
\citep{2009Natur.460..869K,2011MNRAS.413.3075M}.  Recent work has
shown that SN color is also correlated with SN~Ia spectral features,
with SNe possessing faster ejecta velocities having redder colors at
fixed $M_B$ (or equivalently brighter $M_B$ at fixed color) in samples
with very red SNe excluded \citep{2011ApJ...729...55F}.

Under the assumption that any intrinsic SN~Ia color--luminosity
relation has a smaller effective $\beta$ than that from dust (as seems
likely given we observe $\beta<4.1$), our results are qualitatively
consistent with a scenario in which dust extinction modifies this
intrinsic color--luminosity relation. The lowest \mstellar\ host
galaxies are those with the largest specific SFRs, and therefore the
largest dust content. We would therefore expect to find SNe~Ia with a
larger effective $\beta$ (i.e., closer to the true dust value) in
lower \mstellar\ hosts, which is consistent with our observations. In
more massive, passive host galaxies, we are likely observing a $\beta$
closer to the intrinsic value.

\section{Conclusions}
\label{sec:conclusions}

In this paper we have presented the cosmological results for the
Supernova Legacy Survey (SNLS) three-year SN~Ia sample
\citepalias[SNLS3;][]{2010A&A...523A...7G,2011ApJS..192....1C},
combined with other constraints from the literature. Our SN~Ia sample
contains 472 SNe, including 242 from SNLS
\citepalias{2010A&A...523A...7G}, 123 at low redshift, 93 from
SDSS-SN, and 14 from HST \citepalias[see][]{2011ApJS..192....1C}. We
have performed analyses investigating the cosmological parameters
\omatter, \ok, and the dark energy equation-of-state parameter $w$. A
key aspect of our analysis is the inclusion of all identified SN~Ia
systematic uncertainties directly in our cosmological fits
\citepalias{2011ApJS..192....1C}.  The inclusion of these systematic
uncertainties has two key effects.  Foremost, the uncertainties that
we quote on the cosmological parameters reflect the systematic
component.  Second, correlations in brightness, stretch and color
between different SNe due to their being affected by the same
systematic are accounted for during the cosmological fitting stage. We
also correct for recently identified trends between SN~Ia brightness
and host galaxy stellar mass, and account for the effect of systematic
differences from the use of two independent SN~Ia light curve fitters.

Our main results are:

\begin{enumerate}
\item For simple cosmological fits assuming a flat Universe and
  constant $w$, combining the SNLS3 sample with BAO observations and
  the WMAP7 CMB ``shift'' parameters gives
  $\omatter=0.276^{+0.016}_{-0.013}$ and $w=-1.043^{+0.054}_{-0.055}$,
  where the error is statistical only.  When we include all identified
  SN~Ia systematics in the fits, we find
  $\omatter=0.274^{+0.019}_{-0.015}$ and $w=-1.068^{+0.080}_{-0.082}$.

\item In terms of the contribution towards the uncertainty in
  measuring $w$ in the above fits, our systematic and statistical
  uncertainties are approximately equal (5.5\% and 5.2\%
  respectively). However, the systematic uncertainty error budget is
  dominated by the photometric calibration of the SN fluxes, rather
  than uncertainties related to the astrophysics of the SNe
  themselves. Neglecting calibration uncertainties, likely to be
  dramatically reduced in the future, gives a systematic uncertainty
  of $\sim2$\%.

\item When including the SHOES prior on $H_0$, together with the full
  WMAP7 power spectrum and the power spectrum of LRGs in SDSS DR7, we
  find $\omatter=0.269\pm0.015$ and $w=-1.061^{+0.069}_{-0.068}$ using
  the \texttt{CosmoMC} fitter, a 6.5\% measure of $w$.  When we relax
  the assumption of a flat Universe, we find $\omatter=0.271\pm0.015$,
  $\ok=-0.002\pm0.006$, and $w=-1.069^{+0.091}_{-0.092}$. These
  include all SN systematic uncertainties.

\item We consider a simple parameterization of the time variation of
  $w$ as $w(a)=\wo+\wa(1-a)$. Assuming a flat Universe we find
  $\omatter=0.271^{+0.015}_{-0.015}$, $\wo=-0.905^{+0.196}_{-0.196}$
  and $\wa=-0.984^{+1.094}_{-1.097}$. This includes WMAP7, SDSS-DR7,
  and the SHOES $H_0$ prior as external constraints. Our results are
  equivalent to a DETF figure-of-merit of $\sim11$.

\item We investigate astrophysical systematics in our SN~Ia sample by
  breaking it into sub-samples based on SN light curve and host galaxy
  parameters. Cosmologies determined from these SN sub-samples are
  fully consistent. However, we find significant evidence
  ($4.4\sigma$) for a different $\beta$ between low and high stellar
  mass host galaxies (as well as a different $M_B$, which we already
  account for). The effect of this varying $\beta$ on the cosmology
  lies well within the current systematic uncertainty assigned to
  host-galaxy-dependent corrections, but our analysis emphasizes the
  critical need to make host galaxy related corrections when
  determining the cosmological parameters.

\end{enumerate}

When the SNLS3 sample is combined with measurements of large scale
structure, observations of the CMB, and a prior on $H_0$, the
constraints on dark energy presented here are the tightest available,
and directly include all identified SN~Ia systematic uncertainties in
the analysis. All our results are consistent with a flat, $w=-1$
universe.

The primary contributor to the systematic error budget is the
calibration of the SN~Ia fluxes, both placing them on a consistent
system between different SN~Ia surveys, and then interpreting that
system when fitting the SN light curves. The magnitude of the
identified astrophysical systematics is significantly smaller than
calibration related uncertainties. Those that have been uncovered in
the SN Ia population can be adequately controlled using empirical
corrections based on the properties of the SN Ia host galaxies.

The implication is that, if the calibration-related systematics can be
reduced, SNe~Ia are a long way from being systematics limited. In
part, the calibration-related systematics arise from the need to
calibrate the $griz$ SNLS and SDSS filter sets to the $UBVR$ system
used for the majority of the low-redshift SN~Ia data.  This situation
is set to improve considerably as improved low redshift SN~Ia samples
become available
\citep[e.g.,][]{2007PASA...24....1K,2009PASP..121.1395L}.  These new
samples will be directly calibrated to the SNLS system (or vice-versa)
and will eliminate, or at least substantially reduce, the main
systematic uncertainties, allowing the full potential of the SNLS
sample to be unlocked.

\acknowledgments

The SNLS collaboration gratefully acknowledges the assistance of
Pierre Martin and the CFHT Queued Service Observations team.
Jean-Charles Cuillandre and Kanoa Withington were also indispensable
in making possible real-time data reduction at CFHT. We thank Zhiqi
Huang and Adam Riess for useful discussions.

This paper is based in part on observations obtained with
MegaPrime/MegaCam, a joint project of CFHT and CEA/IRFU, at the
Canada-France-Hawaii Telescope (CFHT) which is operated by the
National Research Council (NRC) of Canada, the Institut National des
Sciences de l'Univers of the Centre National de la Recherche
Scientifique (CNRS) of France, and the University of Hawaii. MS
acknowledges support from the Royal Society. Canadian collaboration
members acknowledge support from NSERC and CIAR; French collaboration
members from CNRS/IN2P3, CNRS/INSU and CEA. This work is based in part
on data products produced at the Canadian Astronomy Data Centre as
part of the CFHT Legacy Survey, a collaborative project of NRC and
CNRS.  Based in part on observations obtained at the Gemini
Observatory, which is operated by the Association of Universities for
Research in Astronomy, Inc., under a cooperative agreement with the
NSF on behalf of the Gemini partnership: the National Science
Foundation (United States), the Science and Technology Facilities
Council (United Kingdom), the National Research Council (Canada),
CONICYT (Chile), the Australian Research Council (Australia), CNPq
(Brazil) and CONICET (Argentina).  Based on data from Gemini program
IDs: GS-2003B-Q-8, GN-2003B-Q-9, GS-2004A-Q-11, GN-2004A-Q-19,
GS-2004B-Q-31, GN-2004B-Q-16, GS-2005A-Q-11, GN-2005A-Q-11,
GS-2005B-Q-6, GN-2005B-Q-7, GN-2006A-Q-7, and GN-2006B-Q-10. Based in
part on observations made with ESO Telescopes at the Paranal
Observatory under program IDs 171.A-0486 and 176.A-0589.  Some of the
data presented herein were obtained at the W.M. Keck Observatory,
which is operated as a scientific partnership among the California
Institute of Technology, the University of California and the National
Aeronautics and Space Administration. The Observatory was made
possible by the generous financial support of the W.M. Keck
Foundation. This research has made use of the NASA/IPAC Extragalactic
Database (NED) which is operated by the Jet Propulsion Laboratory,
California Institute of Technology, under contract with the National
Aeronautics and Space Administration.

{\it Facilities:} \facility{CFHT}, \facility{VLT:Antu}, \facility{VLT:Kueyen}, \facility{Gemini:Gillett}, \facility{Gemini:South}, \facility{Keck:I}.


`

\clearpage

\begin{deluxetable}{lcccccc}
\tablecaption{Cosmological results assuming $w=-1$ for the SNLS3 sample plus BAO\tablenotemark{b} and WMAP7\tablenotemark{c}}
\tablecolumns{7}
\tablehead{\colhead{Fit} & \colhead{$\alpha$} & \colhead{$\beta$} & \colhead{$M_B^1$\tablenotemark{a}}  & \colhead{$M_B^2$\tablenotemark{a}} & \colhead{\omatter} & \colhead{\olambda}}

\startdata
\cutinhead{Marginalization fits}
Stat only & $1.451^{+0.123}_{-0.100}$ & $3.165^{+0.105}_{-0.089}$ &  $-19.122$ & $-19.198$ & $0.275^{+0.016}_{-0.013}$ & $0.727^{+0.015}_{-0.013}$\\[4pt]
Stat + sys & $1.428^{+0.121}_{-0.098}$ & $3.263^{+0.121}_{-0.103}$ &  $-19.144$ & $-19.196$ & $0.279^{+0.019}_{-0.015}$ & $0.724^{+0.017}_{-0.016}$\\[4pt]
\cutinhead{$\chi^2$ minimization fits}
Stat only & $1.389^{+0.085}_{-0.083}$ & $3.144^{+0.095}_{-0.092}$ &  $-19.121^{+0.015}_{-0.015}$ & $-19.196^{+0.013}_{-0.013}$ & $0.273^{+0.015}_{-0.014}$ & $0.729^{+0.014}_{-0.014}$\\[4pt]
Stat + sys & $1.368^{+0.086}_{-0.084}$ & $3.182^{+0.102}_{-0.099}$ &  $-19.162^{+0.028}_{-0.029}$ & $-19.206^{+0.024}_{-0.024}$ & $0.274^{+0.017}_{-0.016}$ & $0.732^{+0.016}_{-0.017}$\\[4pt]
\enddata
\tablenotetext{a}{For an $s=1$ and $\col=0$ SN~Ia. Computed from
  ${\mathcal M_{B}}$ ($\S$~\ref{sec:cosm-fits-method}) assuming
  $H_{0}=70$\,km\,s$^{-1}$\,Mpc$^{-1}$. Errors on ${\mathcal M_{B}}$
  are not available in the marginalization (grid) approach as the
  variable is analytically marginalized; the quoted value is an
  estimate only.}  \tablenotetext{b}{\citet{2010MNRAS.401.2148P}}
\tablenotetext{c}{Using the WMAP7 ``shift'' parameter $R$, the
  ``acoustic scale'' $l_a$, and the decoupling redshift $z_{\ast}$
  from \citet{2011ApJS..192...18K}}
\label{tab:cosmofits_wminus1}
\end{deluxetable}

\begin{deluxetable}{lcccccc}
\tablecaption{Cosmological results assuming a flat universe and constant $w$ for the SNLS3 sample plus BAO and WMAP7}
\tablecolumns{7}
\tablehead{\colhead{Fit} & \colhead{$\alpha$\tablenotemark{a}} & \colhead{$\beta$\tablenotemark{a}} & \colhead{$M_B^1$}  & \colhead{$M_B^2$} & \colhead{\omatter} & \colhead{$w$}}
\startdata
\cutinhead{Marginalization fits}
Stat only & $1.450^{+0.112}_{-0.105}$ & $3.164^{+0.096}_{-0.094}$ & $-19.164$\phm{$^{+0.019}_{-0.019}$} & $-19.227$\phm{$^{+0.019}_{-0.019}$} & $0.276^{+0.016}_{-0.013}$ & $-1.043^{+0.054}_{-0.055}$ \\[4pt]
Stat + sys & $1.367^{+0.086}_{-0.084}$ & $3.179^{+0.101}_{-0.099}$ & $-19.175$\phm{$^{+0.019}_{-0.019}$} & $-19.220$\phm{$^{+0.019}_{-0.019}$} & $0.274^{+0.019}_{-0.015}$ & $-1.068^{+0.080}_{-0.082}$ \\[4pt]
\cutinhead{$\chi^2$ minimization fits}
Stat only & $1.395^{+0.085}_{-0.083}$ & $3.148^{+0.095}_{-0.092}$ & $-19.130^{+0.019}_{-0.019}$ & $-19.203^{+0.016}_{-0.016}$ & $0.274^{+0.015}_{-0.014}$ & $-1.039^{+0.052}_{-0.055}$ \\[4pt]
Stat + sys & $1.367^{+0.086}_{-0.084}$ & $3.179^{+0.101}_{-0.099}$ & $-19.155^{+0.027}_{-0.027}$ & $-19.200^{+0.023}_{-0.023}$ & $0.272^{+0.017}_{-0.016}$ & $-1.058^{+0.078}_{-0.082}$ \\
\enddata
\tablenotetext{a}{Note that the values of these nuisance parameters differ very slightly from those from the SN-only fits given in \citetalias{2011ApJS..192....1C} due to small correlations between the cosmological parameters and the nuisance parameters.}
\label{tab:cosmofits}
\end{deluxetable}

\begin{deluxetable}{lccc}
\tablecaption{Detailed summary of systematic uncertainties}
\tablecolumns{4}
\tablehead{\colhead{Source} & \colhead{\omatter} & \colhead{$w$} & \colhead{Relative area\tablenotemark{a}}}
\startdata
Statistical only	            & 0.2763$^{+0.0163}_{-0.0132}$	& $-1.0430^{+0.0543}_{-0.0546}$	 & 1.0\\[2pt]
All systematics		            & 0.2736$^{+0.0186}_{-0.0145}$  	& $-1.0676^{+0.0799}_{-0.0821}$	 & 1.693\\[2pt]
All systematics, except calibration & 0.2756$^{+0.0164}_{-0.0133}$      & $-1.0481^{+0.0573}_{-0.0580}$  & 1.068\\[2pt]
All systematics, except host term   & 0.2738$^{+0.0186}_{-0.0145}$      & $-1.0644^{+0.0790}_{-0.0809}$  & 1.677\\[2pt]
All systematics, fixing $\alpha$, $\beta$\tablenotemark{b} & 0.2656$^{+0.0179}_{-0.0144}$& $-1.1168^{+0.0807}_{-0.0824}$  & 1.641\\[2pt]
\tableline
\sidehead{Contribution of different systematics:}                                                             
 ~~Calibration		            & 0.2750$^{+0.0185}_{-0.0150}$  	& $-1.0581^{+0.0774}_{-0.0791}$	 & 1.614\\[2pt]
 ~~SN Ia model		            & 0.2767$^{+0.0163}_{-0.0132}$  	& $-1.0403^{+0.0543}_{-0.0547}$	 & 1.013\\[2pt]
 ~~Peculiar velocities	            & 0.2761$^{+0.0163}_{-0.0132}$  	& $-1.0452^{+0.0544}_{-0.0548}$	 & 1.002\\[2pt]
 ~~Malmquist bias	            & 0.2758$^{+0.0163}_{-0.0132}$  	& $-1.0474^{+0.0548}_{-0.0553}$	 & 1.014\\[2pt]
 ~~Non SN Ia contamination          & 0.2763$^{+0.0163}_{-0.0132}$	& $-1.0430^{+0.0543}_{-0.0546}$	 & 1.000\\[2pt]
 ~~Milky Way extinction             & 0.2762$^{+0.0164}_{-0.0133}$      & $-1.0441^{+0.0553}_{-0.0557}$	 & 1.023\\[2pt]
 ~~SN redshift evolution            & 0.2763$^{+0.0163}_{-0.0132}$      & $-1.0408^{+0.0544}_{-0.0547}$	 & 1.017\\[2pt]
 ~~Host galaxy term	            & 0.2762$^{+0.0163}_{-0.0132}$      & $-1.0453^{+0.0556}_{-0.0562}$	 & 1.029\\[2pt]
\tableline
\sidehead{Calibration:}                                                             
  ~~Colors of BD 17$^{\circ}$~4708  & 0.2719$^{+0.0170}_{-0.0137}$	& $-1.0720^{+0.0639}_{-0.0639}$	 & 1.239\\[2pt]
  ~~SED of BD 17$^{\circ}$~4708     & 0.2771$^{+0.0170}_{-0.0138}$	& $-1.0390^{+0.0623}_{-0.0630}$  & 1.205\\[2pt]
  ~~SNLS zeropoints	            & 0.2767$^{+0.0168}_{-0.0136}$  	& $-1.0421^{+0.0603}_{-0.0609}$	 & 1.166\\[2pt]
  ~~Low-z zeropoints	            & 0.2753$^{+0.0164}_{-0.0133}$  	& $-1.0527^{+0.0578}_{-0.0586}$	 & 1.078\\[2pt]
  ~~SDSS zeropoints	            & 0.2767$^{+0.0164}_{-0.0133}$  	& $-1.0411^{+0.0544}_{-0.0548}$	 & 1.015\\[2pt]
  ~~SNLS filters	            & 0.2789$^{+0.0170}_{-0.0138}$  	& $-1.0330^{+0.0585}_{-0.0586}$  & 1.136\\[2pt]
  ~~Lowz filters	            & 0.2766$^{+0.0163}_{-0.0132}$  	& $-1.0402^{+0.0547}_{-0.0550}$	 & 1.010\\[2pt]
  ~~SDSS filters	            & 0.2770$^{+0.0164}_{-0.0133}$  	& $-1.0396^{+0.0544}_{-0.0548}$	 & 1.007\\[2pt]
  ~~HST zeropoints	            & 0.2769$^{+0.0164}_{-0.0133}$  	& $-1.0412^{+0.0544}_{-0.0548}$  & 1.007\\[2pt]
  ~~NICMOS nonlinearity	            & 0.2767$^{+0.0164}_{-0.0133}$  	& $-1.0418^{+0.0545}_{-0.0548}$	 & 1.009\\[2pt]
\tableline
\sidehead{SN Ia model (light curve fitter):}                                                        
  ~~SALT2 vs. SiFTO	            & 0.2767$^{+0.0163}_{-0.0132}$  	& $-1.0404^{+0.0543}_{-0.0547}$	 & 1.012\\[2pt]
  ~~Color uncert. model	            & 0.2763$^{+0.0163}_{-0.0132}$  	& $-1.0430^{+0.0543}_{-0.0546}$	 & 1.001\\[2pt]   
\tableline
\sidehead{SN Ia redshift evolution:}                                                             
   ~~$\alpha$		            & 0.2763$^{+0.0163}_{-0.0132}$  	& $-1.0430^{+0.0543}_{-0.0546}$	 & 1.000\\[2pt]
   ~~$\beta$		            & 0.2763$^{+0.0163}_{-0.0132}$  	& $-1.0408^{+0.0544}_{-0.0547}$	 & 1.017\\
\enddata
\tablenotetext{a}{The area of the \omatter--$w$ 68.3\% confidence
  contour relative to a fit considering statistical errors only. The
  contours are computed with the marginalization (grid) approach,
  include BAO and WMAP7 constraints, and assume a flat universe; see
  text for details.}
\tablenotetext{b}{All our cosmological results have $\alpha$ and $\beta$ free in the fits. This entry shows the effect of incorrectly holding them fixed.}
\label{tab:syserrorbudget}
\end{deluxetable}

\begin{deluxetable}{lccccccc}
\tablecaption{Cosmological results\tablenotemark{a} obtained using the \texttt{CosmoMC} fitter with a constant dark energy equation of state}
\tablecolumns{8}
\tabletypesize{\scriptsize}
\tablehead{\colhead{Parameter} & \colhead{WMAP7+SNLS3} & \colhead{WMAP7+DR7} & \colhead{WMAP7+$H0$\tablenotemark{b}} & \colhead{WMAP7+DR7} & \colhead{WMAP7+DR7} & \colhead{WMAP7+$H_0$} & \colhead{WMAP7+DR7}\\ \colhead{} & \colhead{} & \colhead{} & \colhead{} & \colhead{+$H_0$} & \colhead {+SNLS3} & \colhead {+SNLS3} & \colhead{+$H_0$+SNLS3}}
\startdata
\sidehead{Flat, constant $w$:}
~~\omatter & \phs$0.262^{+0.023}_{-0.023}$ & \phs$0.329 ^{+0.034}_{-0.039}$ & \phs$0.246^{+0.020}_{-0.020}$&\phs$0.267 ^{+0.017}_{-0.017}$ & \phs$0.284 ^{+0.019}_{-0.019}$ & \phs$0.250^{+0.017}_{-0.017}$ & \phs$0.269 ^{+0.015}_{-0.015}$\\[4pt]
~~$w$      & $-1.016^{+0.077}_{-0.079}$    & $-0.826 ^{+0.166}_{-0.161}$    & $-1.114^{+0.113}_{-0.113}$   &$-1.110 ^{+0.122}_{-0.120}$    & $-1.021 ^{+0.078}_{-0.079}$    & $-1.037^{+0.068}_{-0.068}$    & $-1.061 ^{+0.069}_{-0.068}$ \\[4pt]
~~$H_0$    & $71.58^{+2.41}_{-2.42}$       & $64.42^{+4.25}_{-4.38}$        & $74.11^{+2.58}_{-2.55}$      &$72.21^{+2.46}_{-2.42}$        & $69.77^{+2.07}_{-2.07}$        & $72.85^{+1.78}_{-1.77}$       & $71.57^{+1.65}_{-1.65}$        \\
\sidehead{Non-flat, constant $w$:}                                                                                         
~~\omatter & \phs$0.259^{+0.050}_{-0.049}$ & \phs$0.312 ^{+0.051}_{-0.051}$ & \nodata                      &\phs$0.253 ^{+0.021}_{-0.021}$ & \phs$0.294 ^{+0.021}_{-0.021}$ & $0.247^{+0.018}_{-0.018}$     & \phs$0.271 ^{+0.015}_{-0.015}$\\[4pt]
~~\ok      & \phs$0.001^{+0.015}_{-0.015}$ & $-0.006^{+0.013}_{-0.012}$     & \nodata                      &$-0.012^{+0.008}_{-0.008}$     & $-0.009^{+0.008}_{-0.008}$     & $0.004^{+0.007}_{-0.007}$     & $-0.002 ^{+0.006}_{-0.006}$\\[4pt]
~~$w$      & $-1.018^{+0.113}_{-0.110}$    & $-1.027 ^{+0.379}_{-0.386}$    & \nodata                      &$-1.445^{+0.298 }_{-0.292}$    & $-1.068^{+0.094}_{-0.095}$     & $-1.001^{+0.092}_{-0.092}$    & $ -1.069 ^{+0.091}_{-0.092}$\\[4pt]
~~$H_0$    & $72.65^{+6.59}_{-6.73}$       & $66.36^{+5.73}_{-5.87}$        & \nodata                      &$73.54^{+2.77}_{-2.79}$        & $67.85^{+2.58}_{-2.57}$        & $73.64^{+2.24}_{-2.24}$       & $71.18^{+1.92}_{-1.87}$     \\
\enddata
\tablenotetext{a}{The values quoted are the expectation values of the
  marginalized distributions, not the best fits, with the 68.3\%
  marginalized values quoted as the errors. All SN systematic
  uncertainties are included. Note that as the non-SN constraints used
  in \texttt{CosmoMC} differ slightly from those used in
  Table~\ref{tab:cosmofits}, the cosmological parameters are
  different. The closest comparison is SNLS3+WMAP7+DR7.}
\tablenotetext{b}{We show only flat universe fits for the WMAP+$H_0$
  combination; the fits were not constraining for non-flat
  cosmologies, with the lower bound on $w$ unconstrained.}
\label{tab:cosmomcfits_constw}
\end{deluxetable}

\begin{deluxetable}{llcc}
\tablecaption{The full set of cosmological parameters obtained with the \texttt{CosmoMC} fitter}
\tablecolumns{4}
\tablehead{\colhead{Class} & \colhead{Parameter} & \colhead{Const. $w$} & \colhead{Const. $w$}\\\colhead{} & \colhead{} & \colhead{flat} & \colhead{non-flat}}
\startdata
 Primary&$     100\Omega_b h^2$ & $ 2.258^{+ 0.054}_{- 0.054}$    & $ 2.265^{+ 0.056}_{- 0.056}$\\[4pt]
        &$            \Omega_c$ & $ 0.1149^{+ 0.0041}_{- 0.0041}$ & $ 0.1145^{+ 0.0047}_{- 0.0047}$\\[4pt]
        &$              \theta$ & $ 1.0398^{+ 0.0026}_{- 0.0027}$ & $ 1.0401^{+ 0.0026}_{- 0.0026}$\\[4pt]
        &$                \tau$ & $ 0.087^{+ 0.006}_{- 0.007}$    & $ 0.088^{+ 0.007}_{- 0.007}$\\[4pt]
        &$            \Omega_k$ & \nodata & $-0.002^{+ 0.006}_{- 0.006}$\\[4pt]
        &$                 w_0$ & $-1.061^{+ 0.069}_{- 0.068}$ & $-1.069^{+ 0.091}_{- 0.092}$\\[4pt]
        &$                 n_s$ & $ 0.969^{+ 0.013}_{- 0.013}$ & $ 0.970^{+ 0.014}_{- 0.013}$\\[4pt]
        &$\log[10^{10} A_{05}]$ & $ 3.095^{+ 0.033}_{- 0.033}$ & $ 3.094^{+ 0.033}_{- 0.033}$\\[4pt]
        &$              \alpha$ & $ 1.451^{+ 0.109}_{- 0.109}$ & $ 1.454^{+ 0.112}_{- 0.111}$\\[4pt]
        &$               \beta$ & $ 3.265^{+ 0.111}_{- 0.111}$ & $ 3.259^{+ 0.111}_{- 0.109}$\\[4pt]
\tableline
 Derived&$\Omega_{\mathrm{DE}}$ & $ 0.731^{+ 0.015}_{- 0.015}$ & $ 0.731^{+ 0.015}_{- 0.015}$\\[4pt]
        &$        \mathrm{Age}$\tablenotemark{a} & $13.71^{+ 0.11}_{- 0.11}$ Gyr  & $13.78^{+ 0.31}_{- 0.31}$ Gyr \\[4pt]
        &$            \Omega_m$ & $ 0.269^{+ 0.015}_{- 0.015}$ & $ 0.271^{+ 0.015}_{- 0.015}$\\[4pt]
        &$            \sigma_8$ & $ 0.850^{+ 0.038}_{- 0.038}$ & $ 0.847^{+ 0.038}_{- 0.038}$\\[4pt]
        &$     z_{\mathrm{re}}$\tablenotemark{b} & $10.55^{+ 1.20}_{- 1.19}$ & $10.55^{+ 1.20}_{- 1.18}$\\[4pt]
        &$                 H_0$ & $71.57^{+ 1.65}_{- 1.65}$ km s$^{-1}$ Mpc$^{-1}$ & $71.18^{+ 1.92}_{- 1.87}$ km s$^{-1}$ Mpc$^{-1}$\\
\enddata
\tablenotetext{a}{The current age of the universe}
\tablenotetext{b}{The redshift at which the reionization fraction is a half}
\label{tab:cosmomc_allparams}
\end{deluxetable}

\begin{deluxetable}{lccc}
\tablecaption{Cosmological results obtained with \texttt{CosmoMC} assuming a variable dark energy equation of state and a flat universe}
\tablecolumns{4}
\tablehead{\colhead{Parameter} & \colhead{WMAP7+DR7} & \colhead{WMAP7+DR7} & \colhead{WMAP7+DR7}\\\colhead{} & \colhead{+SNLS3} & \colhead{+SNLS3+$H_0$}& \colhead{+SNLS3+$H_0$}\\\colhead{} & \colhead{} & \colhead{(stat. only)} & \colhead{(stat+sys)}}
\startdata
\omatter & \phs$0.282 ^{+0.019}_{-0.019}$  & \phs$0.274 ^{+0.014}_{-0.014}$ & \phs$0.271 ^{+0.015}_{-0.015}$\\[4pt]
\wo      & $-0.949 ^{+0.198}_{-0.201}$     & $-0.870 ^{+0.139}_{-0.139}$    & $-0.905 ^{+0.196}_{-0.196}$   \\[4pt]
\wa      & $-0.535 ^{+1.109}_{-1.111}$     & $-0.938 ^{+0.821}_{-0.827}$    & $-0.984 ^{+1.094}_{-1.097}$   \\[4pt]
$H_0$    & $70.26^{+2.40}_{-2.43}$         & $71.38^{+1.40}_{-1.38}$        & $71.99^{+1.68}_{-1.69}$      \\
FoM\tablenotemark{a}      & 10.6 & 21.5 & 11.1\\
\enddata
\tablenotetext{a}{The DETF \citep{2006astro.ph..9591A} figure of merit
  (FoM), implemented here as $1/(\sigma_{w_p}\sigma_{w_a})$; see
  $\S$~\ref{sec:comparison}.}
\label{tab:cosmomcfits_varw}
\end{deluxetable}

\begin{deluxetable}{lcccccc}
\tablecaption{Cosmological results fitting for a different $w$/\omatter\ in each SNLS field}
\tabletypesize{\small}
\tablehead{\colhead{Fit} & \colhead{$\omatter$} & \colhead{$w$}& \colhead{$\alpha$} & \colhead{$\beta$} & \colhead{$\chi^2$} & \colhead{r.m.s.}}
\startdata
Basic fit & $0.272^{+0.017}_{-0.016}$ & $-1.058^{+0.078}_{-0.082}$ & $1.367^{+0.086}_{-0.084}$ & $3.179^{+0.102}_{-0.099}$ & 418.1 & 0.153\\[4pt]
4 field fit\tablenotemark{a} & $0.265^{+0.105}_{-0.133}$,  & $-1.044^{+0.300}_{-0.300}$,   & $1.356^{+0.086}_{-0.084}$ & $3.183^{+0.103}_{-0.101}$ & 414.9 & 0.153\\
&$0.311^{+0.095}_{-0.123}$,&$-1.235^{+0.312}_{-0.358}$,&&\\
&$0.241^{+0.111}_{-0.127}$,&$-0.931^{+0.222}_{-0.245}$,&&\\
&$0.268^{+0.100}_{-0.123}$\phm{,}&$-1.058^{+0.258}_{-0.310}$\phm{,}&&\\
\enddata
\tablenotetext{a}{Fitting for global nuisance parameters and a
  different \omatter\ and $w$ in each of the four SNLS fields. The
  average values of the four \omatter\ and $w$ values are applied to
  non-SNLS SNe. See $\S$~\ref{sec:segr-snls-field} for details.}
\label{tab:fieldvar}
\end{deluxetable}

\begin{deluxetable}{lccccccccc}
\tablecaption{Fits for \omatter\ and $w$ using SN~Ia sub-samples}
\tabletypesize{\scriptsize}
\rotate
\tablecolumns{10}
\tablehead{\colhead{Sample} & \colhead{N} & \colhead{$\alpha$} & \colhead{$\beta$} & \colhead{$M_B^1$}  & \colhead{$M_B^2$} & \colhead{$\sigma_{\mathrm{diff}}$\tablenotemark{a}} & \colhead{\omatter} & \colhead{$w$} & \colhead{r.m.s.}}
\startdata
\sidehead{Statistical:}
All\tablenotemark{b}           & 368 & $1.354\pm0.081$ & $3.097\pm0.094$ & $-19.119\pm0.020$ & $-19.197\pm0.016$ & $5.1$ & $0.281\pm0.015$ & $-1.013\pm0.055$ & 0.139 \\
$s\leq1.0$    & 174 & $1.736\pm0.180$ & $2.837\pm0.141$ & $-19.147\pm0.031$ & $-19.253\pm0.029$ & $4.1$ & $0.280\pm0.016$ & $-1.020\pm0.068$ & 0.135 \\
$s>1.0$       & 194 & $1.998\pm0.251$ & $3.275\pm0.136$ & $-19.054\pm0.034$ & $-19.126\pm0.029$ & $3.3$ & $0.284\pm0.016$ & $-0.994\pm0.070$ & 0.143 \\
$\col\leq0.0$ & 178 & $1.462\pm0.125$ & $3.792\pm0.370$ & $-19.069\pm0.036$ & $-19.126\pm0.028$ & $2.0$ & $0.284\pm0.017$ & $-0.990\pm0.072$ & 0.132 \\
$\col>0.0$    & 190 & $1.267\pm0.120$ & $3.797\pm0.197$ & $-19.173\pm0.031$ & $-19.286\pm0.026$ & $4.4$ & $0.281\pm0.017$ & $-1.012\pm0.072$ & 0.148 \\
$\mstellar\leq10$ & 134 & $1.337\pm0.199$ & $3.575\pm0.169$ & $-19.142\pm0.032$ & \nodata & \nodata & $0.275\pm0.017$ & $-1.064\pm0.090$ & 0.141 \\
$\mstellar>10$    & 234 & $1.319\pm0.087$ & $2.802\pm0.114$ & \nodata & $-19.185\pm0.016$ & \nodata & $0.283\pm0.016$ & $-0.990\pm0.057$ & 0.134 \\
\tableline
\sidehead{Statistical+Systematic:}
ALL           & 368 & $1.341\pm0.082$ & $3.084\pm0.099$ & $-19.128\pm0.028$ & $-19.193\pm0.024$ & $3.0$ & $0.282\pm0.018$ & $-1.004\pm0.084$ & 0.138 \\
$s\leq1.0$    & 174 & $1.753\pm0.185$ & $2.895\pm0.161$ & $-19.142\pm0.040$ & $-19.260\pm0.036$ & $3.4$ & $0.280\pm0.018$ & $-1.022\pm0.093$ & 0.137 \\
$s>1.0$       & 194 & $1.981\pm0.254$ & $3.246\pm0.141$ & $-19.063\pm0.040$ & $-19.133\pm0.037$ & $2.5$ & $0.284\pm0.019$ & $-0.993\pm0.098$ & 0.143 \\
$\col\leq0.0$ & 178 & $1.478\pm0.135$ & $3.981\pm0.416$ & $-19.077\pm0.046$ & $-19.118\pm0.037$ & $1.2$ & $0.281\pm0.020$ & $-1.011\pm0.109$ & 0.135 \\
$\col>0.0$    & 190 & $1.246\pm0.123$ & $3.826\pm0.199$ & $-19.191\pm0.040$ & $-19.297\pm0.038$ & $3.0$ & $0.282\pm0.020$ & $-1.003\pm0.107$ & 0.149 \\
$\mstellar\leq10$ & 134 & $1.320\pm0.202$ & $3.481\pm0.178$ & $-19.158\pm0.043$ & \nodata & \nodata & $0.272\pm0.020$ & $-1.085\pm0.125$ & 0.139 \\
$\mstellar>10$    & 234 & $1.308\pm0.089$ & $2.854\pm0.124$ & \nodata & $-19.177\pm0.024$ & \nodata & $0.282\pm0.018$ & $-0.999\pm0.084$ & 0.134 \\
\tableline
\enddata
\tablenotetext{a}{The significance of the difference between $M_B^1$ and $M_B^2$, including covariances.}
\tablenotetext{b}{``ALL'' means all SNe in the redshift restricted sample described in $\S$~\ref{sec:segr-snprops}.}
\label{tab:cosmofitssubsample}

\end{deluxetable}

\begin{deluxetable}{lccccccccc}
\tabletypesize{\footnotesize}
\tablecaption{Nuisance parameter variation for low and high \mstellar\ host galaxies. The $\sigma$ columns give the significance of the difference in that nuisance parameter between the two host types.}
\tablecolumns{10}
\tablehead{\colhead{Fit} & \colhead{$\alpha^1$} & \colhead{$\alpha^2$} & \colhead{$\sigma_{\mathrm{diff}}$} & \colhead{$\beta^1$} & \colhead{$\beta^2$} & \colhead{$\sigma_{\mathrm{diff}}$} & \colhead{$M_B^1$} & \colhead{$M_B^2$} & \colhead{$\sigma_{\mathrm{diff}}$} }
\startdata
\sidehead{Statistical:}
1$\alpha$,1$\beta$,1$M_B$ & $1.29 \pm 0.08$ & \nodata & \nodata & $3.15 \pm 0.09$ & \nodata & \nodata & $-19.188 \pm 0.015$ & \nodata & \nodata\\
1$\alpha$,1$\beta$,2$M_B$ & $1.39 \pm 0.08$ & \nodata & \nodata & $3.14 \pm 0.09$ & \nodata & \nodata & $-19.130 \pm 0.019$ & $-19.203 \pm   0.016$ & 5.0 \\
1$\alpha$,2$\beta$,1$M_B$ & $1.27 \pm 0.08$ & \nodata & \nodata & $3.70 \pm 0.16$ & $2.78 \pm 0.11$ & 4.7 & $-19.186 \pm 0.015$ & \nodata & \nodata\\
2$\alpha$,1$\beta$,1$M_B$ & $1.10 \pm 0.14$ & $1.34 \pm 0.09$ & 1.5 & $3.16 \pm 0.09$ & \nodata & \nodata & $-19.191 \pm 0.015$ & \nodata & \nodata\\
1$\alpha$,2$\beta$,2$M_B$ & $1.37 \pm 0.08$ & \nodata & \nodata & $3.64 \pm 0.16$ & $2.81 \pm 0.11$ & 4.3 & $-19.130 \pm 0.019$ & $-19.197 \pm   0.015$ & 4.7 \\
2$\alpha$,1$\beta$,2$M_B$ & $1.43 \pm 0.17$ & $1.39 \pm 0.09$ & 0.2 & $3.14 \pm 0.09$ & \nodata & \nodata & $-19.128 \pm 0.021$ & $-19.203 \pm   0.016$ & 4.5 \\
2$\alpha$,2$\beta$,1$M_B$ & $1.00 \pm 0.15$ & $1.34 \pm 0.09$ & 2.1 & $3.74 \pm 0.17$ & $2.79 \pm 0.11$ & 5.0 & $-19.190 \pm 0.015$ & \nodata & \nodata\\
2$\alpha$,2$\beta$,2$M_B$ & $1.33 \pm 0.17$ & $1.38 \pm 0.09$ & 0.3 & $3.64 \pm 0.16$ & $2.81 \pm 0.11$ & 4.3 & $-19.133 \pm 0.021$ & $-19.198 \pm   0.015$ & 4.0 \\
\sidehead{Statistical+Systematic\tablenotemark{a}:}
1$\alpha$,1$\beta$,1$M_B$ & $1.28 \pm 0.08$ & \nodata & \nodata & $3.19 \pm 0.10$ & \nodata & \nodata & $-19.193 \pm 0.022$ & \nodata & \nodata\\
1$\alpha$,1$\beta$,2$M_B$ & $1.39 \pm 0.09$ & \nodata & \nodata & $3.18 \pm 0.10$ & \nodata & \nodata & $-19.130 \pm 0.025$ & $-19.206 \pm   0.022$ & 4.6 \\
1$\alpha$,2$\beta$,1$M_B$ & $1.27 \pm 0.08$ & \nodata & \nodata & $3.75 \pm 0.17$ & $2.80 \pm 0.12$ & 4.5 & $-19.179 \pm 0.021$ & \nodata & \nodata\\
2$\alpha$,1$\beta$,1$M_B$ & $1.08 \pm 0.14$ & $1.33 \pm 0.09$ & 1.6 & $3.21 \pm 0.10$ & \nodata & \nodata & $-19.196 \pm 0.022$ & \nodata & \nodata\\
1$\alpha$,2$\beta$,2$M_B$ & $1.36 \pm 0.08$ & \nodata & \nodata & $3.65 \pm 0.17$ & $2.85 \pm 0.12$ & 4.0 & $-19.127 \pm 0.024$ & $-19.198 \pm   0.021$ & 4.6 \\
2$\alpha$,1$\beta$,2$M_B$ & $1.41 \pm 0.17$ & $1.38 \pm 0.09$ & 0.2 & $3.18 \pm 0.10$ & \nodata & \nodata & $-19.129 \pm 0.026$ & $-19.206 \pm   0.022$ & 4.6 \\
2$\alpha$,2$\beta$,1$M_B$ & $0.98 \pm 0.16$ & $1.33 \pm 0.09$ & 2.1 & $3.78 \pm 0.17$ & $2.81 \pm 0.12$ & 4.7 & $-19.185 \pm 0.021$ & \nodata & \nodata\\
2$\alpha$,2$\beta$,2$M_B$ & $1.31 \pm 0.18$ & $1.37 \pm 0.09$ & 0.3 & $3.65 \pm 0.17$ & $2.85 \pm 0.12$ & 4.1 & $-19.129 \pm 0.025$ & $-19.198 \pm   0.021$ & 4.2 \\

\enddata
\tablenotetext{a}{We exclude the host galaxy systematic term in these fits; see $\S$~\ref{sec:segr-host-galaxy} for details.}
\label{tab:multi_nuisance}
\end{deluxetable}

\begin{deluxetable}{lcccc}
\tablecaption{Cosmological fits including multiple nuisance parameters for low and high \mstellar\ host galaxies}
\tablecolumns{5}
\tablehead{\colhead{Fit} & \colhead{\omatter} & \colhead{$w$} & \colhead{r.m.s.} & \colhead{\chidof}}
\startdata
\sidehead{Statistical:}
1$\alpha$,1$\beta$,1$M_B$ & $0.269\pm0.014$ & $-1.094\pm 0.055$ & 0.153 & 453.6/467 \\
1$\alpha$,1$\beta$,2$M_B$ & $0.274\pm0.014$ & $-1.039\pm 0.054$ & 0.152 & 429.6/466 \\
1$\alpha$,2$\beta$,1$M_B$ & $0.271\pm0.014$ & $-1.072\pm 0.054$ & 0.152 & 430.2/466 \\
2$\alpha$,1$\beta$,1$M_B$ & $0.270\pm0.014$ & $-1.088\pm 0.055$ & 0.152 & 450.8/466 \\
1$\alpha$,2$\beta$,2$M_B$ & $0.274\pm0.014$ & $-1.034\pm 0.053$ & 0.150 & 410.9/465 \\
2$\alpha$,1$\beta$,2$M_B$ & $0.274\pm0.014$ & $-1.039\pm 0.054$ & 0.153 & 429.7/465 \\
2$\alpha$,2$\beta$,1$M_B$ & $0.272\pm0.014$ & $-1.064\pm 0.054$ & 0.151 & 425.7/465 \\
2$\alpha$,2$\beta$,2$M_B$ & $0.274\pm0.014$ & $-1.034\pm 0.053$ & 0.150 & 410.7/464 \\
\sidehead{Statistical+Systematic\tablenotemark{a}:}
1$\alpha$,1$\beta$,1$M_B$ & $0.266\pm0.016$ & $-1.116\pm 0.081$ & 0.154 & 448.4/467 \\
1$\alpha$,1$\beta$,2$M_B$ & $0.273\pm0.016$ & $-1.055\pm 0.079$ & 0.153 & 423.1/466 \\
1$\alpha$,2$\beta$,1$M_B$ & $0.276\pm0.016$ & $-1.037\pm 0.076$ & 0.153 & 425.3/466 \\
2$\alpha$,1$\beta$,1$M_B$ & $0.267\pm0.016$ & $-1.111\pm 0.081$ & 0.153 & 445.9/466 \\
1$\alpha$,2$\beta$,2$M_B$ & $0.274\pm0.016$ & $-1.040\pm 0.075$ & 0.151 & 405.4/465 \\
2$\alpha$,1$\beta$,2$M_B$ & $0.272\pm0.016$ & $-1.054\pm 0.079$ & 0.153 & 423.1/465 \\
2$\alpha$,2$\beta$,1$M_B$ & $0.275\pm0.016$ & $-1.041\pm 0.076$ & 0.152 & 421.1/465 \\
2$\alpha$,2$\beta$,2$M_B$ & $0.274\pm0.016$ & $-1.040\pm 0.075$ & 0.150 & 405.3/464 \\
\enddata
\tablenotetext{a}{We exclude the host galaxy systematic term in these fits; see $\S$~\ref{sec:segr-host-galaxy} for details.}
\label{tab:multi_cosmo}
\end{deluxetable}

\end{document}